\documentclass[aps,prd,twocolumn,amsmath,amssymb,nofootinbib]{revtex4-1}
\usepackage{graphicx}
\usepackage{color}
\usepackage{times}
\usepackage{inputenc}
\usepackage{bm}
\usepackage{ulem}
\usepackage{multirow}
\usepackage{float}
\usepackage{url}
\usepackage{natbib}
\usepackage[colorlinks=true,citecolor=blue,urlcolor=blue,linkcolor=blue]{hyperref}
\begin{document}
\title{Impacts of dark matter on the $f$-mode oscillation of hyperon star}
\author{H. C. Das$^{1,2}$}
\email{harish.d@iopb.res.in}
\author{Ankit Kumar$^{1,2}$}
\author{S. K. Biswal$^{3}$}
\author{S. K. Patra$^{1,2}$} 
\affiliation{\it $^{1}$Institute of Physics, Sachivalaya Marg, Bhubaneswar 751005, India}
\affiliation{\it $^{2}$Homi Bhabha National Institute, Training School Complex,	Anushakti Nagar, Mumbai 400094, India}
\affiliation{\it $^{3}$Department of Engineering Physics, DRIEMS Autonomous Engineering College, Cuttack 754022, India }
\date{\today}
\begin{abstract}
We investigate the $f$-mode oscillation of the dark matter admixed hyperon star within the relativistic Cowling approximation. The macroscopic properties are calculated with the relativistic mean-field equation of states by assuming that the dark matter particles are inside the star. The $f$-mode oscillation frequencies (only for $l=2$) are calculated with four different neutron star equation of states. We also check the effects of hyperons/dark matter and hyperons with dark matter equation of states on the $f$-mode oscillations varying with different astrophysical quantities such as mass ($M$), radius ($R$), compactness ($M/R$), surface red-shift ($Z_s$), average density ($\bar{\rho}$), dimensionless tidal deformability ($\Lambda$) of the neutron star. Significant changes have been seen in the $f$-mode frequencies with and without hyperons/dark matter or hyperons+dark matter. Substantial correlations are observed between canonical frequencies and $\Lambda$ ($f_{1.4}-\Lambda_{1.4}$) and maximum frequencies and canonical $\Lambda$ ( $f_{max}-\Lambda_{1.4}$).
\end{abstract}
\maketitle
\section{Introduction}
\label{sec:intro}
The gravitational waves (GWs) coming from the coalescence of binary neutron star (BNS) merger event decipher enough information that provides strong constraints on the equation of state (EOS) and the internal composition of the neutron star (NS) ~\cite{Abbott_2017, Abbott_2018, Bauswein_2017, Annala_2018, Fattoyev_2018, Radice_2018, Mallik_2018, Most_2018, Tews_2018, Nandi_2019, Capano_2020}. The detected GWs, including the electromagnetic counterparts, open multi-messenger astronomy which helps us to study the properties of compact objects in a different way. Till now, only two BNS mergers were detected, which are GW170817 ~\cite{Abbott_2017} and GW190425 ~\cite{Abbott_2020}. In the future, terrestrial detectors such as LIGO/Virgo/KAGRA may detect more BNS merger events, which can constraint the properties of the compact stars more tightly.  In addition to this, the latest massive pulsar observation PSR J0640+6620 ~\cite{Cromartie_2020} and the simultaneous observation of mass-radius ($M-R$) by {\it Neutron star Interior Composition Explorer (NICER)} put a limit on the $M-R$ profile of the NS ~\cite{Miller_2019, Riley_2019}.

Furthermore, the GWs coming from the oscillation of the compact objects is another way to explore the stars' micro and macroscopic properties. This is because the oscillation frequency mainly depends on the internal structure of the star. In Refs. ~\cite{Anderson_1996, Anderson_1998}, they claimed that one could predict the $M$ and $R$ of the NS with the observation of such frequency as an inverse problem. The formalism to describe the GWs from the oscillation of the NS had been well established ~\cite{Lindblom_1983, Anderson_1996, Anderson_1998}. NS oscillates with different quasi-normal modes, mainly classified according to the restoring force that brings the system to equilibrium positions. Such modes are $f$-mode, $p$-mode, $w$-mode, etc., which provide us various information about the internal structure and EOSs of the NS. Sotani {\it et al.} ~\cite{Sotani_2011} claimed that one could find the signature of hadron-quark phase transitions with the observation of $f$ and $g$-modes frequencies. In Ref. ~\cite{Flores_2014}, they found that if a compact object emits GWs in the range 0--1 kHz, it should be elucidated as hybrid stars, and if the range of the frequency is more than 7 kHz, it should be interpreted as strange stars. Similar type work by Igancio {\it et al.} ~\cite{Sandoval_2018}, proposed that, if the frequencies of $g$-modes are in the limit between 1--1.5 kHz, it should be interpreted as evidence of sharp hadron-quark phase transitions. Hence, we conclude that there is still uncertainty on different mode frequencies because the calculations are model-dependent. Thus, for the first time, we calculate the $f$-mode frequency of the dark matter (DM) admixed NS, including hyperons.

Different observations such as galaxy rotation curves, velocity dispersions, galaxy clusters, gravitational lensing, cosmic microwave background, etc., give sufficient hints about the presence of DM in the Galaxies \footnote{ See  \href{https://en.wikipedia.org/wiki/Dark\_matter}{wikipedia}}. If the DM exists, then it must be accreted inside compact objects like NS. The accretion process may be different from one astrophysical object to other, but the amount of accreted DM particles depend on the nature and age of the compact objects ~\cite{Sandin_2009}. In this case, we take weakly interacting massive particles (WIMPs) as a DM candidate, which are already accreted inside the NS ~\cite{Panotopoulos_2017, Das_2019, Das_2020, Das_2021, DasMNRAS_2021, DasPRD_2021}. After accretion, the DM particles interact with baryons which are discussed in the Sub-Sec. \ref{form:DM}. With the addition of DM, the EOS of the NS becomes softer and $M$, $R$ and dimensionless tidal deformability ($\Lambda$) decreases ~\cite{Quddus_2020, Das_2020, Das_2021, DasMNRAS_2021, DasPRD_2021}. Different phenomena such as the cooling rate become faster for DM admixed NS  ~\cite{Bhat_2019}. If the self-gravitating DM particle having a mass more than the Chandrasekhar limit, it forms a mini black hole and destroys the NS ~\cite{Sandin_2009}. 

It is well known that hyperons appear in the NS interior at densities around $2-3\rho_0$. The presence of the hyperons makes the EOS softer, and as a consequence, the maximum mass is substantially reduced to be incompatible with observation. In addition to DM, we also include hyperons inside the NS. Hyperons appear at the higher density region or core of the NS. The study of hyperons appearances inside the NS is not a brand-new idea. It is fast put forwarded by Cameron {\it et al.} ~\cite{Cameron_1959}. According to Ambartsumyan \& Saakyan ~\cite{Ambartsumyan_1960}, the core of a massive NS consists of an inner hyperon core and an outer nucleon shell. The detailed discussion on hyperons productions and interactions with mesons and their coupling parameters had been studied in Refs. ~\cite{Pais_1966, Millener_1988, NKGfp_1992, NKGPRL_1991, Schaffner_1994, Schaffner_1996, Batty_1997, Schaffner_2000, Harada_2005, Harada_2006, Kohno_2006, Friedman_2007, Weissenborn_2012, *WeissenbornE_2014, Lopes_2014, Bhuyan_2017, Biswal_2019, Biswalaip_2019}. Although different approaches have been tried to solve the issues such as (i) potential depths for each hyperon, (ii) coupling between hyperons-mesons, and (iii) hyperon puzzles, these are still open problems. This is because only some hypernuclei have been discovered in the terrestrial laboratory ~\cite{HAYANO_1989, MARES_1995, Nagae_1998, Saha_2004}. More number of hypernuclei will help us to constraint the coupling parameters by fitting the potential depth. In this study, the hyperons--scalar mesons coupling constants are calculated by fitting with hyperon potential depth, while for hyperons--vector mesons, SU(6) method is used as mentioned in Refs. ~\cite{Lopes_2018, Weissenborn_2012, WeissenbornNPA_2012}. Another efficient approach to fix the hyperons-mesons coupling parameters is SU(3) group theory method as given in Refs. ~\cite{Weissenborn_2012, Tsuyoshi_2013, Lopes_2014, Lopes_2021}. The coupling constants for both scalar and vector mesons are constrained using SU(3) method. One can calculate nuclear matter (NM) and NS properties by using both SU(3) and SU(6) methods and compared between these methods as done in Refs. ~\cite{Weissenborn_2012, Lopes_2014}. We calculate the EOS for DM admixed hyperon star, which is the main ingredient to find the $M$, $R$, $\Lambda$, and $f$-mode frequency. 

We assume the DM particles interact with the baryon octet by the Higgs exchange. In our previous studies ~\cite{Das_2020, Das_2021, DasMNRAS_2021, DasPRD_2021}, we have calculated the NS properties by considering that the DM only interacts with nucleons via Higgs exchange. Hence, the total Lagrangian is the addition of NS and DM. In addition to this, we include hyperons part with the NS Lagrangian (see Sub-Sec. \ref{form:hyp}). Till now, hyperons-mesons and hyperon-DM coupling constants are uncertain. In Ref. ~\cite{Popolo_2020}, they have tried to solve the hyperon puzzle using DM. They have considered different DM masses by changing the DM interaction strength parameter and constrained the whole parameter space with recently observed NS mass. Although they have parametrized the DM mass and interaction strength, they didn't give any satisfactory answers to hyperon-mesons coupling. We hope future direct/indirect DM detection experiments such as DAMA, CDMS, GEDEON, LUX, PANDA, and XENON may constrain the baryon-Higgs form factor. In this case, to solve the $f$-mode frequency for DM admixed hyperon star, we take the values of DM-Higgs coupling, and baryons-Higgs form factors are same as compared to our previous calculations ~\cite{Das_2020, Das_2021, DasMNRAS_2021, DasPRD_2021}. 

In the present study, we calculate the $f$-mode oscillation of the DM admixed hyperon star with different macroscopic observables by employing the relativistic Cowling approximation. The calculations are done for the $l=2$ case (quadrupole). In the future, we may extend our calculations to find different modes frequencies such as $p$, $g$, and $w$-modes with higher values of $l$ in the full general relativity (GR) method as done in Refs.  ~\cite{Lindblom_1983, Anderson_1996, Wen_2019}. The paper is organized as; The formalism for the calculations of EOSs and $M-R$ with the addition of DM+hyperons are given in Sec. \ref{sec:form}. The Sub-Sec. \ref{form:EOS} corresponds to relativistic mean-field (RMF) formalism. Sub-Sec. \ref{form:MR} is devoted to the $M-R$ profile, and the $f$-mode formalism is given in Sub-Sec. \ref{form:fmode}. In Sec. \ref{sec:results}, we consistently present our numerical results. Finally, the summary and conclusions are enumerated in Sec. \ref{sec:summ}
\section{Formalism}	
\label{sec:form}
\subsection{Calculation of EOSs of the NS}
\label{form:EOS}
\subsubsection{\bf Interaction between Baryons-Mesons and Mesons-Mesons}
\label{form:hyp}
In this section, we adopt the RMF model to calculate the NS properties. In RMF model, the nucleons interact with each other by exchanging different mesons such as isoscalar-scalar ($\sigma$), isoscalar-vector ($\omega$), isovector-vector ($\rho$), isovector-scalar ($\delta$) ~\cite{Miller_1972,Serot_1986,Frun_1997,Reinhard_1986} and the hyperon-hyperon interaction is mediated by the two additional strange mesons, strange scalar ($\sigma^*$) and strange vector ($\phi$) ~\cite{Scaffner_1996,Biswal_2019,Pradhan_2021}. Hence, the Lagrangian of the system includes the interaction from mesons-nucleons, mesons-hyperons and their self and cross-couplings. The extended RMF (E-RMF) model include the possible significant interactions between mesons up to the fourth-order ~\cite{Kumar_2017,Kumar_2018, Das_2020, Kumar_2020, Das_2021, DasBig_2020, DasMNRAS_2021, DasPRD_2021}. We also add the leptons ($e^-$ and $\mu^-$) contribution to the E-RMF Lagrangian because they are required to maintain both $\beta$-equilibrium and charge neutrality conditions which provide stability to the  NS. Therefore, the E-RMF Lagrangian of the NS system is given as ~\cite{Biswal_2019, Kumar_2020, Das_2021, DasMNRAS_2021}
\begin{eqnarray}
{\cal L}_{NS}&=& \sum_{B}\bar{\psi}_{_B}\Bigg\{i\gamma^{\mu}\partial_{\mu}-m_{_B}+g_{\sigma_B}\sigma-g_{\omega _B}\gamma_{\mu}\omega^{\mu}
\nonumber \\
&-&\frac{1}{2}g_{\rho_B}\gamma_{\mu}\vec{\tau}_{{_B}}\!\cdot\!\vec{\rho}^{\,\mu}+g_{\delta_B}\vec{\tau}_{_B}\!\cdot\!\vec{\delta}\Bigg\}\psi_{_B}+\frac{1}{2}\partial_{\mu}\sigma\,\partial^{\mu}\sigma
\nonumber \\
&-&m_{\sigma}^{2}\sigma^2\Big(\frac{1}{2}+\frac{\kappa_3}{3!}\frac{g_{\sigma_B} \sigma}{m_{_B}}+\frac{\kappa_4}{4!}\frac{g_{\sigma_B}^2\sigma^2}{m_{_B}^2}\Bigg)-\frac{1}{4}F_{\mu\nu}F^{\mu\nu}
\nonumber \\
&+&\frac{1}{2}m_{\omega}^{2}\omega_{\mu}\omega^{\mu}\Big(1+\eta_1\frac{g_{\sigma_B}\sigma}{m_{_B}}+\frac{\eta_2}{2}\frac{g_{\sigma_B}^2\sigma^2}{m_{_B}^2}\Big)-\frac{1}{4}\vec R_{\mu\nu}\!\cdot\!\vec R^{\mu\nu}	
\nonumber\\
&+&\frac{\zeta_0}{4!}g_{\omega_B}^2(\omega_{\mu}\omega^{\mu})^2+\frac{1}{2}m_{\rho}^{2}(\vec\rho_{\mu}\!\cdot\!\vec\rho\ ^{\mu})\Big(1+\eta_{\rho}\frac{g_{\sigma_B}\sigma}{m_{_B}}\Big)
\nonumber\\
&+&\Lambda_{\omega}g_{\omega_B}^2g_{\rho_B}^2(\omega_\mu\omega^\mu\times\vec\rho_\mu\!\cdot\!\vec\rho\ ^\mu)^2+\frac{1}{2}\Big(\partial_{\mu}\vec\delta\,\partial^{\mu}\vec\delta-m_{\delta}^{2}\vec\delta^{\,2}\Big),
\nonumber \\
&+&{\cal L}_{YY}+{\cal L}_l,
\label{eq:NSlag}
\end{eqnarray}
where 
\begin{eqnarray}
{\cal L}_{YY}&=&\sum_Y  \bar{\psi}_{_Y}  (g_{\sigma^*_Y} \sigma^*-g_{\phi_ Y}\gamma_{\mu}\phi^{\mu})\psi_{_Y}+\frac{1}{2}m_{\phi}^2 \phi_{\mu}\phi^{\mu}
\nonumber\\
&-&\frac{1}{4} \phi_{\mu \nu}\phi^{\mu \nu}  +\frac{1}{2}  (\partial_{\mu} \sigma^* \partial^{\mu}\sigma^* - m_{\sigma^*}^2 {\sigma^*}^2), \\
\mathrm{and} \nonumber \\
{\cal L}_l &=& \sum_{l}\bar\psi_{l}\Big(i\gamma^{\mu}\partial_\mu-m_l\Big)\psi_l.
\end{eqnarray}
The $m_{_B}$ represent the masses of the baryons. $m_\sigma$, $m_\omega$, $m_\rho$, $m_\delta$, $m_{\sigma^*}$ and $m_\phi$ are the masses and $g_{\sigma_B}$, $g_{\omega_B}$, $g_{\rho_B}$, $g_{\delta_B}$, $g_{\sigma^*_B}$ and $g_{\phi_B}$ are the coupling constants for the $\sigma$, $\omega$, $\rho$, $\delta$, $\sigma^*$ and $\phi$ mesons respectively. $\kappa_3$ (or $\kappa_4$) and $\zeta_0$ are the self-interacting coupling constants of the $\sigma$ and $\omega$ mesons respectively. $\eta_1$, $\eta_2$, $\eta_\rho$ and $\Lambda_\omega$ are the cross-coupling constants as indicated in the Lagrangian. The quantities $F^{\mu\nu}$, $\vec R^{\mu\nu}$, and $\Phi^{\mu\nu}$ are the field strength tensors for the $\omega$, $\rho$ and $\phi$ mesons respectively, defined as $F^{\mu\nu}$ = $\partial^\mu\omega^\nu-\partial^\nu\omega^\mu$, $\vec R^{\mu\nu}$ = $\partial^\mu\vec\rho^{\,\nu}-\partial^\nu\vec\rho^{\,\mu}$, and $\Phi^{\mu\nu}=\partial^\mu\phi^{\,\nu}-\partial^\nu\phi^{\,\mu}$. The $\vec\tau_{_B}$ is the isospin operator, which carries the isospin component of the baryons. $m_l$ represent the masses of leptons. $Y$ represents different hyperons such as $\Lambda$, $\Sigma^{+,-,0}$ and $\Xi^{+,-}$.
\begin{table*}
\caption{The ratio of masses of different mesons with nucleon mass ($M=939$ MeV), nucleons-mesons coupling constants, self and cross-couplings between mesons are tabulated for considered parameter sets NL3 ~\cite{Lalazissis_1997}, IOPB-I ~\cite{Kumar_2018}, FSUGarnet ~\cite{Chen_2014} and G3 ~\cite{Kumar_2017}. All coupling constants are dimensionless.}
\label{tab:table1}
\scalebox{1.1}{
\begin{tabular}{llllllllllllllll}
\hline \hline
Model &
$\hspace{0.15 cm}\frac{m_\sigma}{M}$ &
$\hspace{0.15 cm}\frac{m_\omega}{M}$ &
$\hspace{0.15 cm}\frac{m_\rho}{M}$ &
$\hspace{0.15 cm}\frac{m_\delta}{M}$ &
$\hspace{0.15 cm}\frac{g_{\sigma{_N}}}{4\pi}$ &
$\hspace{0.15 cm}\frac{g_\omega{_N}}{4\pi}$ &
$\hspace{0.15 cm}\frac{g_{\rho{_N}}}{4\pi}$ &
$\hspace{0.15 cm}\frac{g_{\delta{_N}}}{4\pi}$ &
$\hspace{0.15 cm}\kappa_3$ &
$\hspace{0.15 cm}\kappa_4$ &
$\hspace{0.15 cm}\zeta_0$ &
$\hspace{0.15 cm}\eta_1$ &
$\hspace{0.15 cm}\eta_2$ &
$\hspace{0.15 cm}\eta_\rho$ &
$\hspace{0.15 cm}\Lambda_\omega$ \\ \hline
NL3 & 0.541 & 0.833 & 0.812 & 0.000 & 0.813 & 1.024 & 0.712 & 0.000 & 1.465 & -5.688 & 0.000 & 0.000 & 0.000 & 0.000 & 0.000 \\ \hline
IOPB-I & 0.53 & 0.833 & 0.812 & 0.000 & 0.827 & 1.062 &       0.885 & 0.000 & 1.496 & -2.932 & 3.103 & 0.000 & 0.000 & 0.000 &0.024 \\ \hline
FSUGarnet & 0.529 & 0.833 & 0.812 & 0.000 & 0.837 & 1.091 & 1.105 & 0.000 & 1.368 & -1.397 & 4.410 & 0.000 & 0.000 & 0.000 & 0.043 \\ \hline
G3 & 0.559 & 0.832 & 0.820 & 1.043 & 0.782 & 0.923 & 0.962 & 0.160  & 2.606 & +1.694 & 1.010 & 0.424 & 0.114 & 0.645 & 0.038 \\ \hline \hline
\end{tabular}}
\end{table*}

The equations of motion correspond to different mesons are calculated in the Refs. ~\cite{Kumar_2018,Biswal_2019,Kumar_2020} within the mean-field approximations. By applying the stress tensor technique, the energy density and pressure of the system can be calculated as ~\cite{Kumar_2018, Biswal_2019, Kumar_2020}:
\begin{eqnarray}
{\cal E}_{NS}& = & \sum_B \frac{\gamma_{_B}}{2\pi^2}\int_{0}^{k_{F_B}} k^2\ dk \sqrt{k^2+m^{*2}_{_B}}+n_{_B} g_{\omega_B}\omega_0
\nonumber\\
&+&\frac{n_{3B}}{2}g_{\rho{_B}}\rho_0-\frac{1}{3!}\zeta_{0}{g_{\omega_B}^2}\omega_0^4-\Lambda_{\omega}g_{\rho_B}^2g_{\omega_B}^2\rho_{03}^2\omega_0^2
\nonumber\\
&+&m_{\sigma}^2{\sigma_0}^2\Bigg(\frac{1}{2}+\frac{\kappa_{3}}{3!}\frac{g_{\sigma_B}\sigma_0}{m_{_B}}+\frac{\kappa_4}{4!}\frac{g_{\sigma_B}^2\sigma_0^2}{m_{_B}^2}\Bigg)
\nonumber\\
&-&\frac{1}{2}m_{\omega}^2\omega_0^2\Bigg(1+\eta_{1}\frac{g_{\sigma_B}\sigma_0}{m_{_B}}+\frac{\eta_{2}}{2}\frac{g_{\sigma_B}^2\sigma_0^2}{m_{_B}^2}\Bigg)
\nonumber\\
&-&\frac{1}{2}\Bigg(1+\frac{\eta_{\rho}g_{\sigma_B}\sigma_0}{m_{_B}}\Bigg)m_{\rho}^2\rho_{03}^{2}
+\frac{1}{2}m_{\delta}^2 \delta_0^{2}+\frac{1}{2}m_\phi^2\phi_0^2
\nonumber\\
&+&\frac{1}{2}m_{\sigma^*}^2\sigma_0^{*2}+\sum_l\frac{\gamma_{_l}}{2\pi^2}\int_0^{k_{F{_l}}} dk \ \sqrt{k^2+m_l^2},
\label{eq:ENS}
\end{eqnarray}
and
\begin{eqnarray}
P_{NS}& = & \sum_B \frac{\gamma_{_B}}{6\pi^2}\int_{0}^{k_{F_B}} \frac{k^4\ dk}{\sqrt{k^2+m^{*2}_{_B}}} +\frac{1}{3!}\zeta_{0}{g_{\omega_B}^2}\omega_0^4
\nonumber\\
&-&m_{\sigma}^2{\sigma_0}^2\Bigg(\frac{1}{2}+\frac{\kappa_{3}}{3!}\frac{g_{\sigma_B}\sigma_0}{m_{_B}}+\frac{\kappa_4}{4!}\frac{g_{\sigma_B}^2\sigma_0^2}{m_{_B}^2}\Bigg)-\frac{1}{2}m_{\delta}^2 \delta_0^{2}
\nonumber\\
&+&\frac{1}{2}m_{\omega}^2\omega_0^2\Bigg(1+\eta_{1}\frac{g_{\sigma_B}\sigma_0}{m_{_B}}+\frac{\eta_{2}}{2}\frac{g_{\sigma_B}^2\sigma_0^2}{m_{_B}^2}\Bigg)+\frac{1}{2}m_{\sigma^*}^2\sigma_0^{*2}
\nonumber\\
&+&\frac{1}{2}\Bigg(1+\frac{\eta_{\rho}g_{\sigma_B}\sigma_0}{m_{_B}}\Bigg)m_{\rho}^2\rho_{03}^{2}
+\Lambda_{\omega}g_{\rho_B}^2g_{\omega_B}^2\rho_{03}^2\omega_0^2
\nonumber\\
&+&\frac{1}{2}m_\phi^2\phi_0^2+\sum_l\frac{\gamma_{_l}}{6\pi^2}\int_0^{k_{F{_l}}} \frac{k^2 dk}{\sqrt{k^2+m_l^2}},
\label{eq:PNS}
\end{eqnarray}
where $\gamma_{_B}$ and $\gamma_{l}$ are the spin degeneracy factor for baryons and leptons, respectively. $k_{F_B}$ and $k_{F_l}$ are the baryons and leptons Fermi momentum, respectively. $m^*_{_B}$ is the effective masses of the baryons, which is written by 
\begin{eqnarray}
m^*_B &=& m_B-g_{\sigma_B}\sigma_0-g_{\delta_B}\tau_B\delta_0-g_{\sigma^*_B}\sigma^*.
\label{eq:effm}
\end{eqnarray}
The last term in Eq. (\ref{eq:effm}) doesn't contribute to the $m^*_B$. This is because the value of $g_{\sigma^*_B}$ and $g_{\phi_B}$is zero for the nucleons. \\

\noindent
\textit{\bf Coupling constants for Nucleons-Mesons and Mesons-Mesons:-}\\

To calculate the EOS for the NM and NS systems, one must know the coupling constants. Many RMF parameter sets have been developed, including different types of interaction between nucleons-mesons and mesons-mesons (both self and cross). Therefore, to study how the coupling parameters affect both NM and NS properties, we take four different types of parameter sets such as NL3 ~\cite{Lalazissis_1997}, IOPB-I ~\cite{Kumar_2018}, FSUGarnet ~\cite{Chen_2014}, and G3 ~\cite{Kumar_2017}. We tabulate the masses of different mesons, coupling constants in Table \ref{tab:table1} for these parameter sets. The NM properties correspond to four parameter sets are also reported in Table 3 of the Ref. ~\cite{Kumar_2018}.  
From Table 3 of the Ref. ~\cite{Kumar_2018}, it is noticed that all three parameter sets reproduce the NM properties very well except NL3. For example, the value of incompressibility, $K=$ 271.38 MeV, is higher than other sets. This is because NL3 is a stiff EOS as compared to G3, which is a softer EOS.

By spanning soft to stiff EOSs, one can have a better knowledge of the DM impacts on the NS properties. Although the incompressibility of NL3 is quite high as compared to other considered sets, the predicted properties of the finite nuclei by NL3 are as good as the other sets. In case of G3, it is one of the latest extended RMF (E-RMF) parameter set, which includes almost all significant possible interaction terms between mesons and nucleons (either self or cross) \cite{Kumar_2017}. Also, G3 predicts NS mass around $2\ M_\odot$, and it satisfies NICER data as well. By these forces, we calculate the $f$-mode oscillations of the NS. The EOSs for these sets are shown in Fig. \ref{fig:EOS} to get a clear picture of their variation with energy density.\\

\noindent
\textit{\bf Coupling constants for Hyperons-Mesons}:-\\

Till now, we don't know exactly how the hyperons interact with each other as compared to the nucleons-nucleons interaction ~\cite{NKGPRC_2001}, because of the limitation of the hypernuclei data. Therefore, fixing the couplings between hyperons-mesons is quite difficult. In general, the SU(6) symmetry group is used to fix the couplings between hyperons and vector mesons, while the couplings with scalar mesons are fixed through the hyperon potential depth ~\cite{Pais_1966, Lopes_2014, Biswal_2019, Biswalaip_2019}. The potential depth of the $\Lambda$-hyperon ($U_\Lambda$) is known to be -28 MeV ~\cite{Millener_1988,NKGPRL_1991,Schaffner_1994,Batty_1997, Torres_2017, Lopes_2018} but the $\Sigma$ and $\Xi$ potentials ($U_\Sigma$, $U_\Xi$) provide a large uncertainty and not even the signs of the potentials are well defined ~\cite{Dover_1984, Schaffner_1994}. But some hyper-nuclear experiments show that the $U_\Sigma$ and $U_\Xi$ potential depth close to +30 MeV and -18 MeV respectively ~\cite{Schaffner_2000,Harada_2005, Harada_2006, Kohno_2006, Friedman_2007}.

The hyperon potential depth is defined as ~\cite{NKGb_1997}
\begin{equation}
U_Y = -g_{\sigma{_N}}x_{\sigma{_Y}}\sigma_0+g_{\omega {_N}}x_{\omega{_Y}}\omega_0,
\label{eq:uYN}
\end{equation}
where $x_{\sigma{_Y}}$ and $x_{\omega{_Y}}$ is defined as $g_{\sigma{_Y}}/g_{\sigma{_N}}$ and $g_{\omega{_Y}}/g_{\omega{_N}}$ respectively. For symmetric NM the quantities $g_{\sigma{_Y}}/g_{\sigma{_N}}$ and $g_{\omega{_Y}}/g_{\omega{_N}}$ are given in Refs. ~\cite{NKGb_1997, Biswal_2019}
\begin{eqnarray}
&&g_{\sigma{_N}}\sigma_0=M-M^*=M\Big(1-\frac{M^*}{M}\Big),  \nonumber \\
&&
g_{\omega{_N}}\omega_0=\Big(\frac{g_{\omega{_N}}}{m_\omega}\Big)^2\rho_0,
\label{eq:NKGY}
\end{eqnarray}
Rewriting Eq. (\ref{eq:uYN}) by inserting Eq. (\ref{eq:NKGY}), we get
\begin{eqnarray}
U_Y = M\Big(\frac{M^*}{M}-1\Big)x_{\sigma{_Y}}+\Big(\frac{g_{\omega{_N}}}{m_\omega}\Big)^2\rho_0x_{\omega{_Y}}.
\end{eqnarray}
In this case, we choose the values of $U_\Lambda =-28$ MeV, $U_\Sigma=+30$ MeV and $U_\Xi = -18$ MeV. To get the desire potential one has to choose the values of $x_{\sigma{_Y}}$ and $x_{\omega{_Y}}$ for a fix parameter set. For example NL3 case, we fix,  $x_{\sigma{_\Lambda}}=0.8$, and the value of $x_{\omega{_\Lambda}}$ is found to be 0.8982. Similarly for IOPB-I set, we fix, $x_{\sigma{_\Lambda}}$ = 0.8 which gives $x_{\omega{_\Lambda}}=0.8338$. For $\Sigma$ and $\Xi$-hyperons the values are  $x_{\sigma_\Sigma}=0.7$, $x_{\omega_\Sigma}=0.8932$, $x_{\sigma_\Xi}=0.8$, and $x_{\omega_\Xi}=0.8639$ for IOPB-I set. Hence, one can get different combinations to fit the potential depth. The hyperon interactions with $\rho$-meson are fitted according to SU(6) symmetry method ~\cite{Dover_1984, Schaffner_1994} and their couplings with hyperons are given as: $x_{\rho{_\Lambda}}=0.0$, $x_{\rho{_\Sigma}}=2.0$ and $x_{\rho{_\Xi}}=1.0$. 

The couplings constants for hyperons and $\sigma^*$-meson are $x_{\sigma^*\Lambda}=0.69$, $x_{\sigma^*\Sigma}=0.69$, and $x_{\sigma^*\Xi}=1.25$. Similarly, for hyperons and $\phi$-mesons coupling constants are $x_{\phi\Lambda}=-\frac{\sqrt{2}}{3}g_{\omega_N}$, $x_{\phi\Sigma}=-\frac{\sqrt{2}}{3}g_{\omega_N}$, and $x_{\phi\Xi}=-\frac{2\sqrt{2}}{3}g_{\omega_N}$ respectively for $\Lambda$, $\Sigma$ and $\Xi$-hyperons ~\cite{Biswal_2019}. In this calculation we neglects the $\sigma^*$ and $\phi$ mesons-nucleons interaction for the numerical simplicity.
\subsubsection{\bf Interaction between Baryons-Dark Matter}
\label{form:DM}
When NS evolves in the Universe, the DM particles are accreted inside it due to its huge gravitational potential and immense baryonic density. In this work, we assume that the DM particles are already inside the NS. With this assumption, we model the interaction Lagrangian for DM and baryons. In this case, the Neutralino is the DM particle having a mass of 200 GeV as considered in our earlier works~~\cite{Das_2020, Das_2021}.

The DM particles are interacted with baryons by exchanging standard model (SM) Higgs. The interacting Lagrangian is given in the form ~\cite{Panotopoulos_2017, Das_2019, Quddus_2020, Das_2020, Das_2021, DasMNRAS_2021, DasPRD_2021}:
\begin{eqnarray}
{\cal{L}}_{DM}&=&  \bar \chi \Big[ i \gamma^\mu \partial_\mu - M_\chi + y h \Big] \chi +  \frac{1}{2}\partial_\mu h \partial^\mu h 
\nonumber\\
&-& \frac{1}{2} M_h^2 h^2 +\sum_B f_{_B} \frac{m_{_B}}{v} \bar{\psi}_{_B} h \psi_{_B} , 
\label{eq:ldm}
\end{eqnarray}
$\psi_{_B}$ and $\chi$ are the baryons and DM wave functions respectively. The parameters $y$ is DM-Higgs coupling, $f_{_B}$ is the baryons-Higgs form factors and $v$ is the vacuum expectation value of Higgs field. The values of $y$ and $v$ are 0.07 and 246 GeV respectively taken from the Refs. ~\cite{Das_2020, Das_2021}. The value of $f_{_B} (=$ 0.35) for all baryons taken to be same, because we don't know the form factor for hyperons except nucleons with Higgs. 

We calculate the spin-independent scattering cross-section of the nucleons with DM using the relation ~\cite{Cline_2013}
\begin{eqnarray}
\sigma_{SI}=\frac{y^2f^2M_{nucl.}^2}{4\pi}\frac{\mu_r}{v^2M_h^2},
\end{eqnarray}
where $\mu_r$ is the reduced mass. The calculated cross-sections for $M_\chi=200$ GeV is found to be $9.70\times10^{-46}$ cm$^{2}$, which is well consistent with XENON-1T ~\cite{Xenon1T_2016}, PandaX-II ~\cite{PandaX_2016}, PandaX-4T \cite{Meng_2021} and LUX ~\cite{LUX_2017} within 90\% confidence level. The LHC had also given a limit on the WIMP-nucleon scattering cross-section in the range from $10^{-40}$ to $10^{-50}$ cm$^{2}$ ~\cite{Djouadi_2012}. For hyperons cases, we find the $\sigma_{SI}$ are 1.37, 1.56, 1.91 $\times$ $10^{-45}$ cm$^2$ for $\Lambda$, $\Sigma$ and $\Xi$ respectively. Thus our model also satisfies the LHC limit. Therefore in the present calculations, we constrained the value of $y$ from both the direct detection experiments and the LHC results.

Nucleon-Higgs form factor ($f$) had been calculated in Ref. ~\cite{Djouadi_2012} using the implication of both lattice QCD ~\cite{Czarnecki_2010} and MILC results ~\cite{MILC_2009} whose value is $0.33_{-0.07}^{+0.30}$ ~\cite{Aad_2015}. The taken value of $f$ (= 0.35 ) in this calculation lies in the region. Thus, we also constrain the values of $f$ with available data. 

The Euler Lagrange equation of motion for DM particle ($\chi$) and Higgs boson ($h$) can be derived from Lagrangian in Eq. (\ref{eq:ldm}) as,
\begin{eqnarray}
&&\bigg(i\gamma^{\mu}\partial_{\mu}-M_{\chi}+yh \bigg)\chi {}=0, \nonumber \\
&&\partial_{\mu}\partial^{\mu}h+M_{h}^2h=y\bar{\chi}\chi+\sum_B\frac{f_{_B} m_{_B}}{v}\bar{\psi_{_B}}\psi_{_B},
\label{eq:motion}
\end{eqnarray}
respectively. Applying RMF approximation, we get \cite{Das_2019},
\begin{eqnarray}
&&h_0=\frac{y\langle\bar{\chi}\chi\rangle+\sum_B f\frac{m_{_B}}{v}\langle\bar{\psi_{_B}}\psi_{_B}\rangle}{M_h^2},\nonumber \\
&& \bigg(i\gamma^{\mu}\partial_{\mu}-M^{\star}_{\chi} \bigg)\chi {}=0,
\end{eqnarray}
where $M_\chi^{\star}$ is the dark matter effective mass can be given as,  
\begin{equation}
M_{\chi}^{\star} {}=M_{\chi}-yh_0. 
\label{eq:m_chi_star}
\end{equation}
The dark matter scalar density ($\rho_s^{DM}$)
is
\begin{equation}
\rho_s^{DM}=\langle\bar{\chi}\chi\rangle =  \frac{\gamma}{2 \pi^2}\int_0^{k_f^{DM}} dk \  \frac{M_\chi^\star}{\sqrt{M_\chi^\star{^2}+ k^2}},
\label{eq:dm_density}
\end{equation}
where $k_f^{DM}$ is the Fermi momentum for dark matter. $\gamma$ is the spin degeneracy factor which has value 2 for neutron and proton individually.

Assuming the average number density of nucleons ($n_b$) is $10^3$ times larger than the average dark matter density ($n_{DM}$), which implies the ratio  of the dark matter and the neutron star mass to be $\sim \frac{1}{6}$ \cite{Panotopoulos_2017}. The nuclear saturation density $n_0 \sim 0.16$ fm$^{-3}$, therefore, the DM number density becomes $n_{DM} \sim 10^{-3}n_0\sim 0.16\times 10^{-3}$ fm$^{-3}$. Using the $n_{DM}$, the $k_{f}^{DM}$ is obtained from the equation $k_f^{DM}=(3\pi^2 n_{DM})^{1/3}$. Hence the value of $k_{f}^{DM}$ is $\sim 0.033$ GeV. Therefore, in our case, we vary the DM momenta from 0 to 0.06 GeV. For IOPB-I+DM3 \footnote{DM3 means DM Fermi momentum $k_f^{DM}=$ 0.03 GeV.} case, the predicted mass and radius is consistent with Cromartie {\it et al.} \cite{Cromartie_2020} and NICER \cite{Miller_2019, Riley_2019, Miller_2021} data respectively (see Fig. \ref{fig:mr}). Hence, we take IOPB-I EOS with DM momentum 0.03 GeV in this calculation.

The energy density (${\cal{E}}_{DM}$) and pressure ($P_{DM}$) for NS with DM can be obtained by solving the Eq. (\ref{eq:ldm})
\begin{eqnarray}
{\cal{E}}_{DM}& = & \frac{1}{\pi^2}\int_0^{k_f^{DM}} k^2 \ dk \sqrt{k^2 + (M_\chi^\star)^2 } +\frac{1}{2}M_h^2 h_0^2 ,
\label{eq:edm}
\end{eqnarray}
and
\begin{eqnarray}
P_{DM}& = & \frac{1}{3\pi^2}\int_0^{k_f^{DM}} \frac{ k^4 \ dk} {\sqrt{k^2 + (M_\chi^\star)^2}} - \frac{1}{2}M_h^2 h_0^2 ,
\label{eq:pdm}
\end{eqnarray} 
$M_h$ is the mass of the Higgs equal to 125 GeV, and $h_0$ is the Higgs field calculated by applying the mean-field approximation ~\cite{Das_2019}. The contribution of the Higgs field in both energy density and pressure is minimal.
\begin{figure}
\centering
\includegraphics[width=0.52\textwidth]{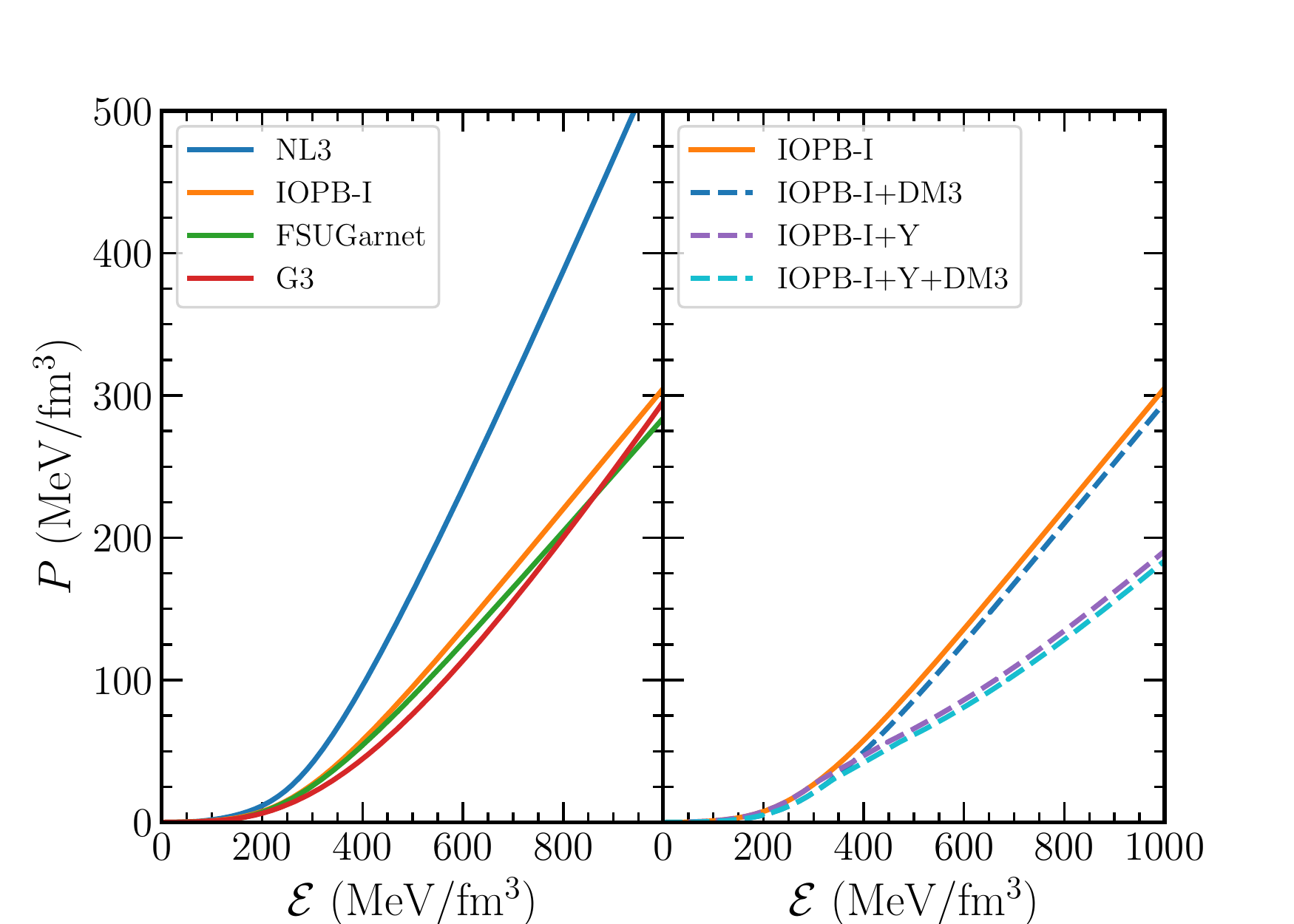}
\caption{{\it Left:} EOSs for different RMF parameter sets. {\it Right}: EOSs are for IOPB-I with yellow line (only nucleons), IOPB-I+DM3 with blue dashed line (nucleons+DM3), IOPB-I+Y with purple dashed line (nucleons+hyperons), IOPB-I+Y+DM with dashed cyan line (nucleons+hyperons+DM3). }
\label{fig:EOS}
\end{figure}	

Therefore, the total energy density (${\cal E}$) and pressure ($P$) for the DM admixed hyperonic NS are as follow:
\begin{eqnarray}
{\cal{E}} =   {\cal{E}}_{NS}+ {\cal{E}}_{DM}, \ {\rm and }\ P =   P_{NS}+ P_{DM}.
\label{eq:EOS}
\end{eqnarray}

In Fig. \ref{fig:EOS}, we plot P with ${\cal E}$ for four considered parameter sets. It is clear from the figure that the NL3 is the stiffest EOS as compared to others, while G3 is the softer EOS. On the right side of the plot, we show the EOS for IOPB-I parameter set for nucleons (IOPB-I), nucleons with DM (IOPB-I+DM3), nucleons with hyperons (IOPB-I+Y), and nucleons with both hyperons and DM (IOPB-I+Y+DM3). The EOS becomes softer with the addition of DM/hyperons or DM+hyperons, but the softness is more for DM+hyperons (dashed cyan line) than others, which is clearly visible in the figure. 

In the presence of hyperons/DM particles, the EOS of neutron star becomes softer. This is because every system would like to minimize its energy. As the density increases, the Fermi momenta or the Fermi energy increases. Because the nucleons are fermions and need to place them in a higher orbit with increasing density. Also, it is well known that the density is proportional to the cube of the Fermi momenta. At sufficiently high density, the energy of the nucleon increases, which is the rest mass energy plus the kinetic energy or $E=\sqrt{k_f^2+M^2}$. When this energy exceeds the mass of the hyperons/kaons/DM, the nucleons decay to these particles. In other words, it is economical energy-wise for the system to have the hyperons/kaons/DM in the lower energy states rather than nucleons at higher Fermi energy at higher density, though the hyperons are heavier than the nucleons. In other words, the nucleons are replaced by hyperons/kaons/DM depending on the system's density. As a result, a fraction of the gravitational mass is converted to kinetic energy and decreases the gravitational mass. With these EOSs, we calculate the $M-R$ profiles and the $f$-mode frequencies of the NS.

\subsection{NS Macroscopic Properties}
\label{form:MR}
The metric for a spherically symmetric and non-rotating relativistic NS is expressed as 
\begin{equation}
ds^2=-e^{2\Phi (r)}dt^2+e^{2\Lambda (r)}dr^2+r^2 d\theta^2+r^2\sin^2{\theta} d\phi^2, \label{eq:metric}
\end{equation}
where $\Phi(r)$ and $\Lambda(r)$ are the metric functions. The spherical symmetric solutions are given by Tolman-Oppenheimer-Volkoff (TOV) coupled equations ~\cite{TOV1, TOV2} as follows: 
\begin{eqnarray}
\frac{dm(r)}{dr}&=&4\pi r^2 {\cal E}(r),
\nonumber\\
\frac{dP(r)}{dr}&=&-\Big[P(r)+{\cal E}(r)\Big] \frac{d\Phi}{dr},
\nonumber \\
\frac{d\Phi (r)}{dr}&=&\frac{m(r)+4\pi r^3P(r)}{r[r-2m(r)]},
\end{eqnarray}
where $m(r)$ represents the enclosing mass at some radius $r$. The hydrostatic equations are solved for a given central density which gives mass and radius of the NS. 
\begin{figure}
\centering
\includegraphics[width=0.52\textwidth]{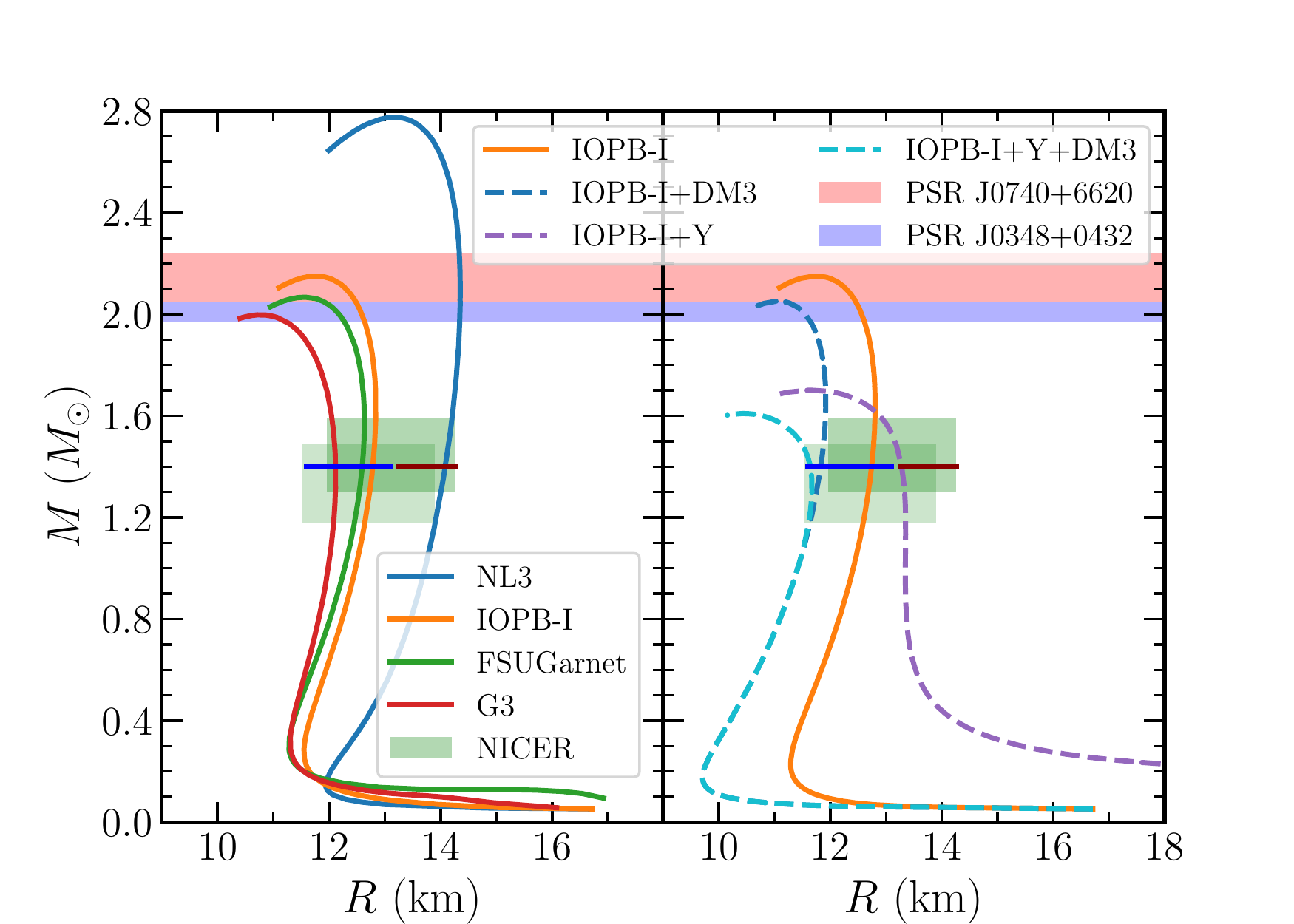}
\caption{{\it Left:} $M$--$R$ relations for four different RMF parameter sets. {\it Right}: $M$--$R$ relations for IOPB-I, IOPB-I+DM3, IOPB-I+Y and IOPB-I+Y+DM3 . The red and blue band represents the masses of the massive pulsars observed by Cromartie {\it et al.} ~\cite{Cromartie_2020} and Antoniadis {\it et al.} ~\cite{Antoniadis_2013}. The old NICER results are shown with two green boxes from two different analyses ~\cite{Miller_2019, Riley_2019}. The new NICER result is shown for canonical radius (blue horizontal line) given by  Miller {\it et al.} ~\cite{Miller_2021}. Reed {\it et al.} radius range is also shown with dark red line for canonical star \cite{Reed_2021}.}
\label{fig:mr}
\end{figure}

We plot the $M-R$ relations for the EOSs on the left side of Fig. \ref{fig:mr}. We observe that the NL3 EOS predicts the maximum mass around 2.77 $M_\odot$ and radius $\sim$14 km. Since it is a stiffer EOS compared to others, its maximum mass is more than the other three sets. We put maximum mass constraints measured from the different pulsars such as PSR \ J0740+6620 ~\cite{Cromartie_2020}, and PSR J0348+0432 ~\cite{Antoniadis_2013}. Except for NL3, the other three EOSs can support maximum NS mass 2.0 $M_{\odot}$, which are compatible with the precise measured NS masses by Cromartie {\it et al.} ($2.14_{-0.09}^{+0.10}\ M_\odot$ with 68.3\% credibility interval) ~\cite{Cromartie_2020} and Antoniadis {\it et al.} ($2.01\pm0.04 \ M_\odot$) ~\cite{Antoniadis_2013}. The simultaneous measurement of mass and radius by the NICER give the $M = 1.44_{-0.14}^{+0.15}\ M_\odot$ ($M = 1.34_{-0.16}^{+0.15}\ M_\odot$) and $R = 13.02_{-1.06}^{+1.24}$ \ km ($R = 12.71_{-1.19}^{+1.14}$ \ km) respectively from the analysis of PSR J0030+0451 by Miller {\it et al.} (Riley {\it et al.}) ~\cite{Miller_2019, Riley_2019}. We called it as old NICER data. We put the old NICER data with two green boxes from two different analyses ~\cite{Miller_2019, Riley_2019}. The predicted radius by all the EOSs are compatible with old NICER data as shown in Fig. \ref{fig:mr}. On the left side of Fig. \ref{fig:mr}, we noticed that IOPB-I and IOPB-I+DM3 could able to produce the maximum mass constraints. The maximum mass and radius correspond to IOPB-I EOS are 2.15 $M_\odot$ and 11.75 km respectively. The maximum masses and radii are (2.05 $M_\odot$, 11.02 \ km), (1.70 $M_\odot$, 11.39\ km), and (1.61 $M_\odot$, 10.40\ km) for IOPB-I+DM3, IOPB-I+Y, and IOPB-I+Y+DM3 respectively. These EOSs satisfy the old NICER data. Recently, Miller {\it et al.} put another radius constraints on both for the canonical ($R_{1.4} = 12.45 \pm 0.65$ \ km) and maximum NS ($R_{2.08} = 12.35 \pm 0.75$ \ km) from the NICER and X-ray Multi-Mirror (XMM) Newton data are termed as new NICER data ~\cite{Miller_2021}. We depicted the new NICER data as a horizontal blue line in Fig. \ref{fig:mr}. All considered EOSs satisfy the new NICER data except NL3 and IOPB-I+Y sets.

Recently, the PREX-2 experiment has given the updated neutron skin thickness of $^{208}$Pb as $R_{\rm skin}=0.284\pm0.071$ fm, within an $1\%$ error \cite{Adhikari_2021}. Based on this data, Reed {\it et al.} inferred the symmetry energy and slope parameter as $J=38.1\pm4.7$ MeV and $L=106\pm37$ MeV respectively, with the help of a limited relativistic mean-field forces (using PREX-2 data) \cite{Reed_2021}. These values of J and L are larger as compared to the old PREX data (PREX-I). The latest $J$ and $L$ can be reproduced mostly from the stiff equation of state. That means the PREX-2 result allows a stiff equation of state. Reed {\it et al.} also predicted the radius for canonical NS as $13.25<R_{1.4}<14.26$ km. In this case, only NL3 satisfies the Reed {\it et al.} data as shown in Fig. \ref{fig:mr}. Recently, Miller {\it et al.} also gave new NICER constraints both for canonical and maximum mass NS from the X-ray study of PSR J0030+0451 ~\cite{Miller_2021}. The improved radius estimate for canonical star is $11.8<R_{1.4}<13.1$ km. This new NICER constraint allows a narrow radius range as compared to old NICER data ($ 11.52<R_{1.4}<14.26$ km). From the radii constraint, we find that the new NICER data allows a narrow radius range contrary to a large range of PREX-2 and the old NICER data, leaving us an inconclusive determination of the NS radius.
\subsection{Calculation of $f$-mode oscillation of the NS}
\label{form:fmode}
NS oscillates when it is disturbed by an external/internal event. Hence, it emits gravitational waves with different modes of frequencies. The most important modes are fundamental mode ($f$), first and second pressure modes ($p$), and the first gravitational mode ($g$) ~\cite{Anderson_1996, Sandoval_2018, Pradhan_2021}. Almost all the energy of the NS is emitted as GWs radiations with these modes. To study different modes of oscillations, one has to solve perturbed fluid equations in the vicinity of GR. Throne and Campollataro ~\cite{Throne_1967} first calculated the non-radial oscillation of NS in the framework of general relativity. Lindblom and Detweiler gave the first integrated numerical solution of the NS ~\cite{Lindblom_1983}. To solve the non-radial oscillations, one can also use the Cowling approximation, which is simpler than the Lindblom and Detweiler method. In Cowling approximations, the metric perturbations are neglected ~\cite{Cowling_1941}. The obtained frequencies within Cowling approximations differ by 10-30\% as compared to full linearized equations of GR. 

In this work, we want to calculate the non-radial oscillations of the NS using the Cowling approximations for a DM admixed hyperon star. To find different oscillations mode  frequencies, one has to solve the following coupled differential equations ~\cite{Sotani_2011, Flores_2014, Sandoval_2018, Pradhan_2021} 
\begin{eqnarray}
\frac{d W(r)}{dr}&=&\frac{d {\cal E}}{dP}\left[\omega^2r^2e^{\Lambda (r)-2\Phi (r)}V(r)
+\frac{d \Phi(r)}{dr} W (r)\right] \nonumber \\
&&
-l(l+1)e^{\Lambda (r)}V (r) \nonumber \\
\frac{d V(r)}{dr} &=& 2\frac{d\Phi (r)}{dr} V (r)-\frac{1}{r^2}e^{\Lambda (r)}W (r).
\label{eq:wv}
\end{eqnarray}
The functions $V (r)$ and $W (r)$ along with frequency $\omega$, characterize the Lagrange displacement vector  ($\eta$) associate to perturbed fluid,
\begin{eqnarray}
    \eta=\frac{1}{r^2}\Big(e^{-\Lambda (r)}W (r),-V (r)\partial_{\theta},
    -\frac{V(r)}{ \sin^{2}{\theta}}\  \partial _{\phi}\Big)Y_{lm},
\end{eqnarray}
where $Y_{lm}$ is the spherical harmonic which is function of $\theta$ and $\phi$. Solution of Eq. (\ref{eq:wv}) with the fixed background metric Eq. (\ref{eq:metric}) near origin will behave as:
\begin{equation}
    W (r)=Br^{l+1}, \ V (r)=-\frac{B}{l} r^l,
\label{eq:bc1}
\end{equation}
where $B$ is an arbitrary constant. At the surface of the star, the perturbation pressure must vanish which provides another boundary condition as follows: 
\begin{equation} 
    \omega^2 e^{\Lambda (R)-2\Phi (R)}V (R)+\frac{1}{R^2}\frac{d\Phi (r)}{dr}\Big|_{r=R}W (R)=0.
\label{eq:bc2}
\end{equation}
After solving Eq. (\ref{eq:wv}) with two boundary conditions (Eqs. (\ref{eq:bc1}) \& (\ref{eq:bc2})), one can get eigenfrequencies of the NS. In this work, we solve the oscillations equations Eq. (\ref{eq:wv}) by using the shooting method with some initial guess for $\omega^2$. The equations are integrated from the center to the surface and trying to match the surface boundary conditions. After each integration, the initial guess of $\omega^2$ is corrected through the Secant method to get the desire precision which improves the initial guess.
\begin{figure}
\centering
\includegraphics[width=0.52\textwidth]{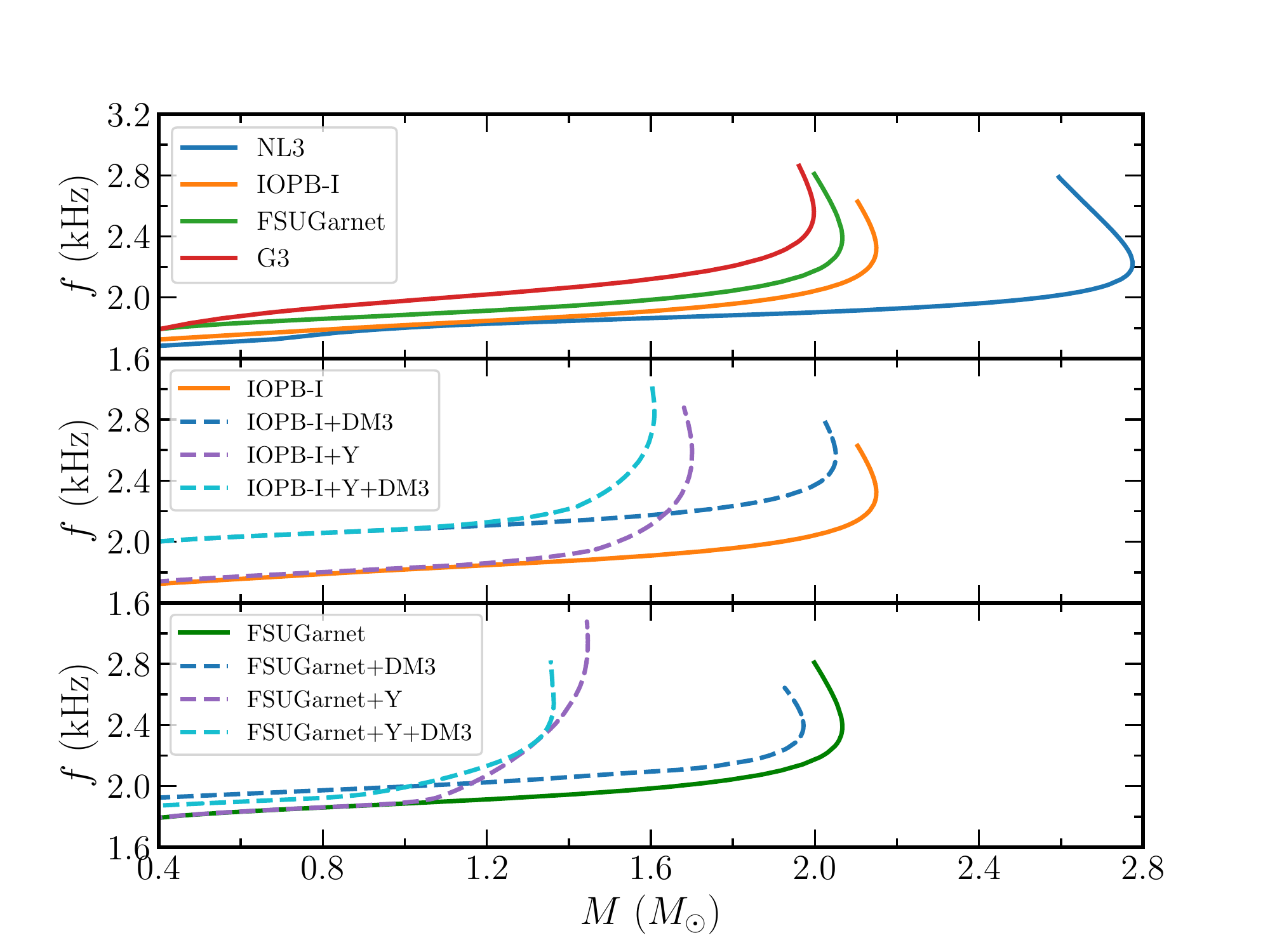}
\caption{{\it Upper:} $f$-mode frequencies as a function of mass for four different RMF parameter sets. {\it Middle}: $f$-mode frequencies for IOPB-I set with DM3/hyperons and hyperons+DM3. {\it Lower}: Same as middle one but for FSUGarnet.}
\label{fig:fm2}
\end{figure}	
\section{Results}
\label{sec:results}
In this Sec. we calculate the $f$-mode oscillation frequencies as the functions of $M$, compactness ($C$), average density ($\bar{\rho}$), surface red-shift ($Z_s$), and $\Lambda$ using the RMF EOSs for DM admixed hyperon star. The EOSs correspond to nucleons, and nucleons+DM3+Y are given in Sub. Sec. \ref{form:EOS}. The formalism for the calculation of $f$-mode oscillation frequencies is given in Sub. Sec. \ref{form:fmode}.
\subsection{Calculation of $f$-mode frequency as functions of different observables}
The $f$-mode frequencies (only for $l=2$) with different RMF EOSs are shown in Fig. \ref{fig:fm2}. In addition to this, the $f$-mode frequencies for IOPB+DM3, IOPB+Y, and IOPB-I+Y+DM are also shown in the middle panel of the figure. To see the parametric dependence of either DM/Y or DM+Y, we repeat the calculations with FSUGarnet and shown in the lower panel of Fig. 3. 

The $f$-mode frequency corresponds to maximum mass ($f_{max}$) for the four EOSs are shown in the upper panel of Fig. 3. The $f_{max}$ are 2.16, 2.32, 2.38, and 2.55 kHz for NL3, IOPB-I, FSUGarnet, and G3, respectively. The $f_{max}$ is maximum for G3 and minimum for NL3, and these are in-between for IOPB-I, and FSUGarnet sets. This is because G3 is softer, and NL3 is the stiffest EOS in our calculations.  In the middle panel of Fig. \ref{fig:fm2}, we observe that there is a negligible change of $f$-mode frequencies up to 1.3 $M_\odot$ for the IOPB-I case. There is no change in the $f$-mode frequencies up to $1 \ M_\odot$ for FSUGarnet as shown in the lower panel of Fig. \ref{fig:fm2}. The change in the $f$-mode frequencies is seen mainly at the core part. This is due to the presence of hyperons and DM particles at the dense region of the NS, which generally occurs in the central region. It is well known that the EOS is model-dependent, mostly at the core region. Also, the appearance of hyperons/DM is not possible in the lower density region. The change in $f-$mode frequency due to hyperon/DM particle needed a minimum mass of the neutron star, which is seen in Ref. \cite{Flores_2020}. Therefore, the change in the $f$-mode frequency varies with forces. 

The $f$-mode frequencies differ considerably with force parametrizations. The values of $f_{max}$ are 2.32, 2.57, 2.58, and 2.85 kHz for IOPB-I, IOPB-I+DM3, IOPB-I+Y, and IOPB-I+Y+DM3, respectively. The EOSs are softer with the addition of either DM3/hyperons or DM3+hyperons compared to the original IOPB-I EOS. Therefore, the $f_{max}$ value is more for IOPB-I+Y+DM3 as compared to others. We tabulated the value of $f_{max}$ and $f_{1.4}$ for four different RMF forces with different NS observables in Table \ref{tab:table4}. The mass variation of frequency changes within interval 1.75-2.55 kHz for $l=2$, which is almost consistent with Pradhan {\it et al.} ~\cite{Pradhan_2021} for the considered parameter sets. For hyperons/DM3 and hyperons+DM3 cases, the frequency range within interval 1.8-2.85 kHz. This range is a little higher as compared with no hyperons/DM case. 
\begin{figure}
\centering
\includegraphics[width=0.52\textwidth]{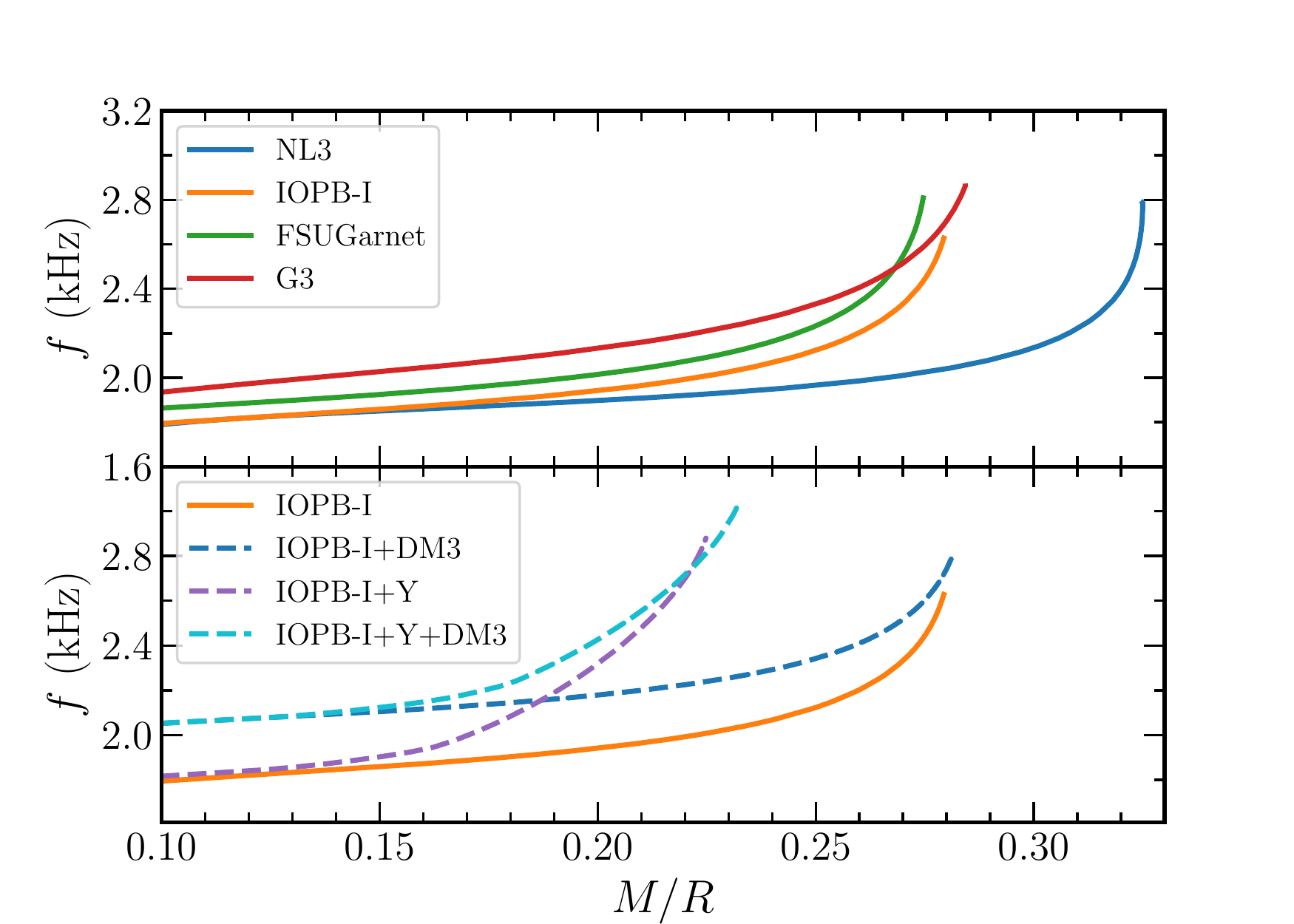}
\caption{{\it Upper:} $f$-mode frequencies as a function of compactness are shown for four different RMF parameter sets. {\it Lower}: $f$-mode frequencies for IOPB-I set with DM3/hyperons and hyperons+DM3.}
\label{fig:fc2}
\end{figure}

The compactness of a star is defined as ($C=M/R$), where $M$ and $R$ are the mass and radius of the star ~\cite{NKGb_1997, schaffner-bielich_2020}. Like $C$, another significant quantity is surface red-shift $Z_s$ is defined as ~\cite{NKGb_1997}
\begin{equation}
    Z_s =\frac{1}{\sqrt{1-\frac{2M}{R}}} -1 = \frac{1}{\sqrt{1-2C}}-1.
\end{equation}
If we find the value of $Z_s$, one can constraints the mass and radius of the star. Till now, only one value of $Z_s=0.35$ is reported in the Ref. ~\cite{Cottam_2002} from the analysis of stacked bursts in the low-mass x-ray binary EXO 0748-676, which is also discarded by the subsequent observation ~\cite{Cottam_2008}.

\begin{figure}
\centering
\includegraphics[width=0.52\textwidth]{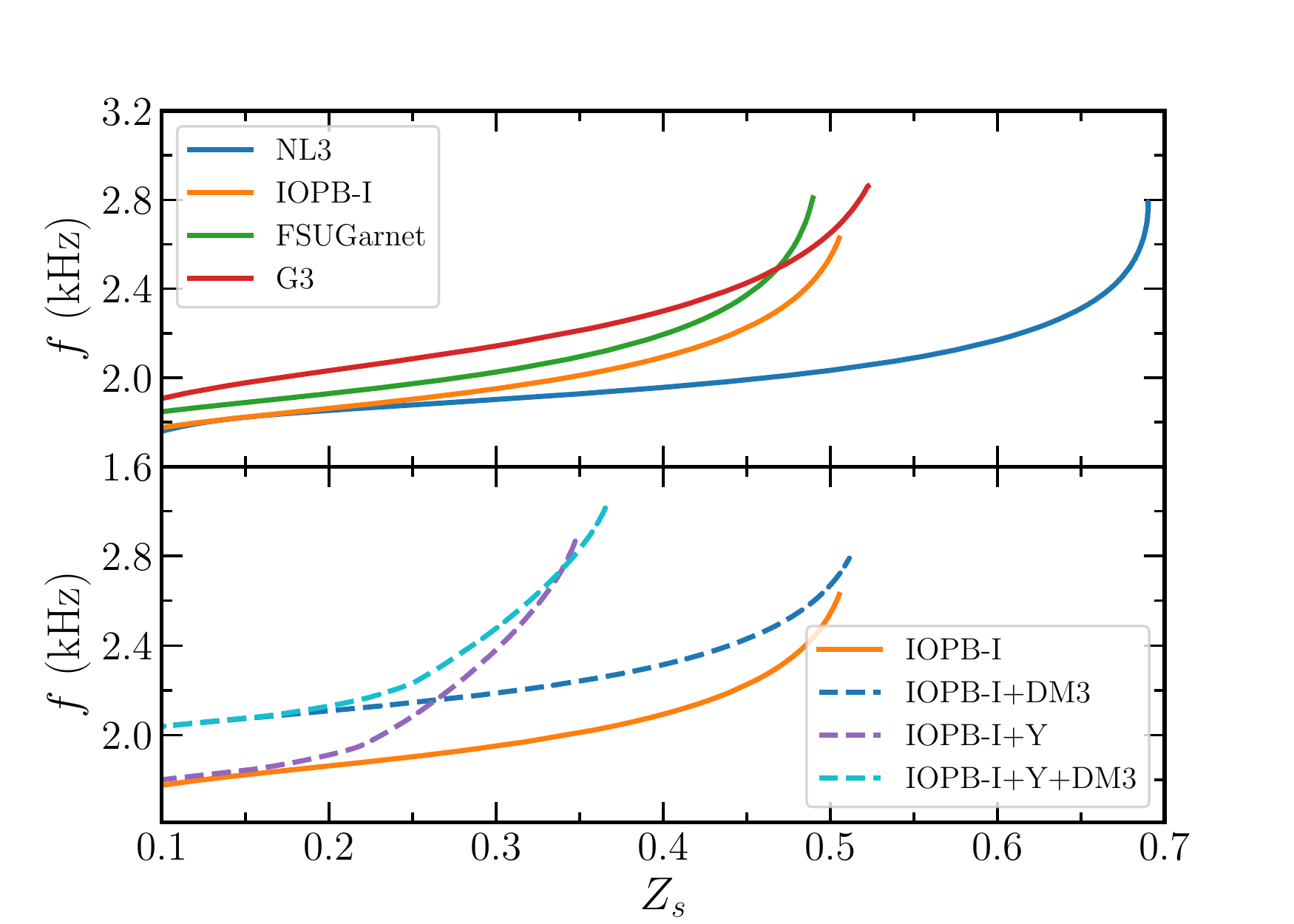}
\caption{Same as Fig. \ref{fig:fc2}, but for $f$ as a function of $Z_s$.}
\label{fig:fr2}
\end{figure}

We calculate $C$ and $Z_s$, which are shown in Figs. \ref{fig:fc2} and \ref{fig:fr2} with $f$-mode frequencies. We observed that the variation of $f$-mode frequencies both for $C$ and $Z_s$ looks almost identical. This is because the $Z_s$ is the function of stellar compactness. The numerical values of $C$ and $Z_s$ correspond to both canonical and maximum mass NS are given in Table \ref{tab:table4}. In our case, we find the range of $C_{max}$ and $Z_{s_{max}}$ are 0.22-0.31 and 0.33-0.62 respectively for the  considered EOSs. The values of $C_{max}$ and $Z_{s_{max}}$ decreases as compared to only nucleonic EOSs.

The average density of a star is defined as $\bar{\rho}= \sqrt{\bar{M}/\bar{R^3}}$, where $\bar{M} = \frac{M}{1.4\ M_\odot} $ and $\bar{R}=\frac{R}{10 \ \mathrm{km}}$. We plot the $f$-mode frequencies with the variation of $\bar{\rho}$. On top of that, we plot some empirical fit relations (see Table \ref{tab:table3}) from previous calculation ~\cite{Anderson_1998, Benhar_2004, Pradhan_2021} including our fitting results.  
\subsection{Fitting formula} 
The relation between $f$-mode frequency as a function of average density was first calculated by Andersson and Kokkotas (AK) ~\cite{Anderson_1996}. They got an empirical relation by using some polytropic EOSs as follow:
\begin{equation}
    f(\mathrm{kHz})\approx 0.17+2.30\sqrt{\left(\frac{M}{1.4 \ M_\odot}\right)\left(\frac{10 \ \mathrm{km }}{R}\right)^3}.
    \label{eq:AK1}
\end{equation}
This empirical relations is again modified by their subsequent paper using some realistic EOSs ~\cite{Anderson_1998}, which is given as 
\begin{equation}
    f(\mathrm{kHz})\approx 0.78+1.635\sqrt{\frac{\bar{M}}{\bar{R^3}}}.
     \label{eq:AK2}
\end{equation}
\begin{figure}
\centering
\includegraphics[width=0.52\textwidth]{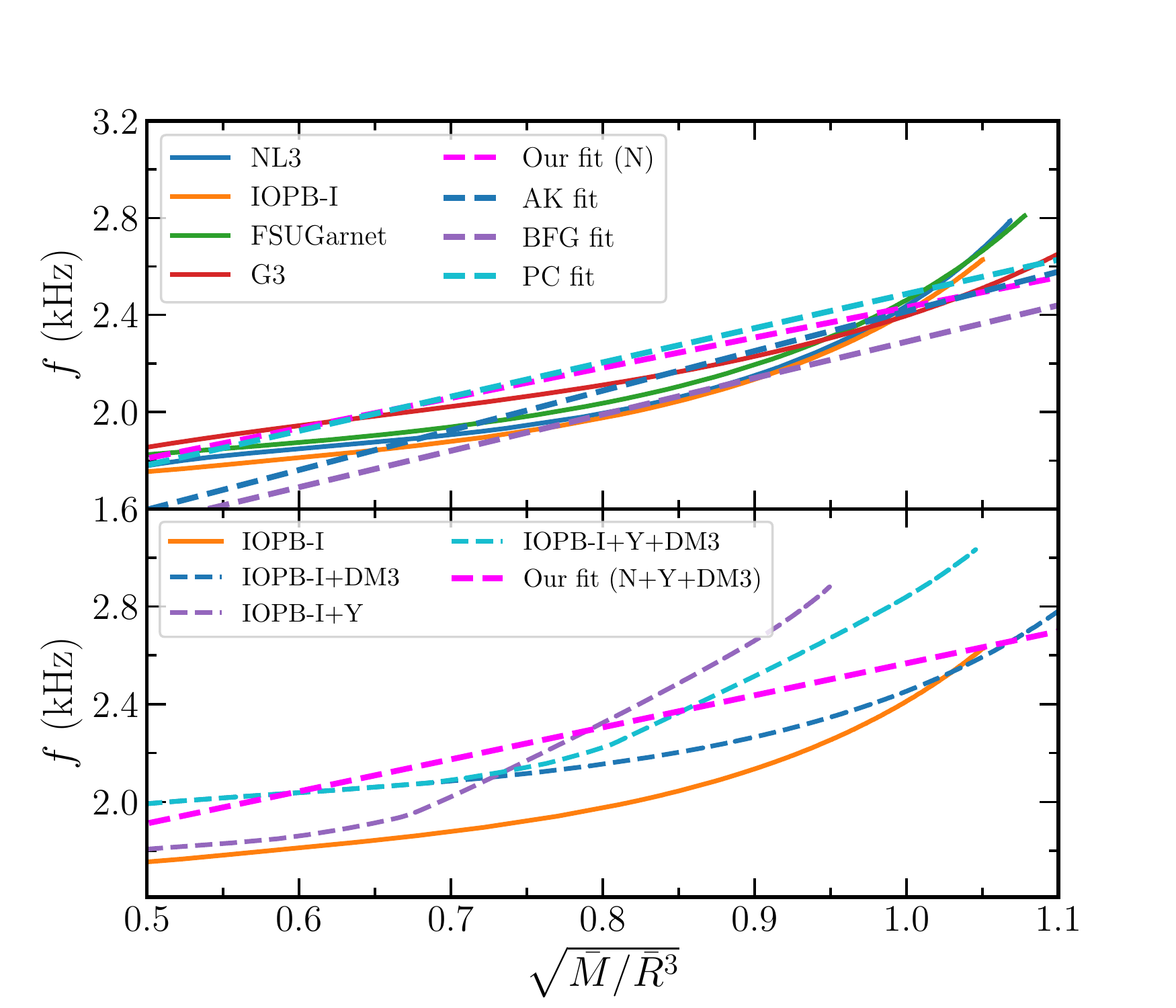}
\caption{{\it Upper:} $f$-mode frequency as a function of average density for four parameter sets. The dashed line with different colours are the fitted relations taken from different analyses including ours. {\it Lower:} The same relations but for nucleonic EOSs along with DM3/Y and DM3+Y.}
\label{fig:fd2}
\end{figure}
\begin{figure}[t]
	\centering
	\includegraphics[width=0.52\textwidth]{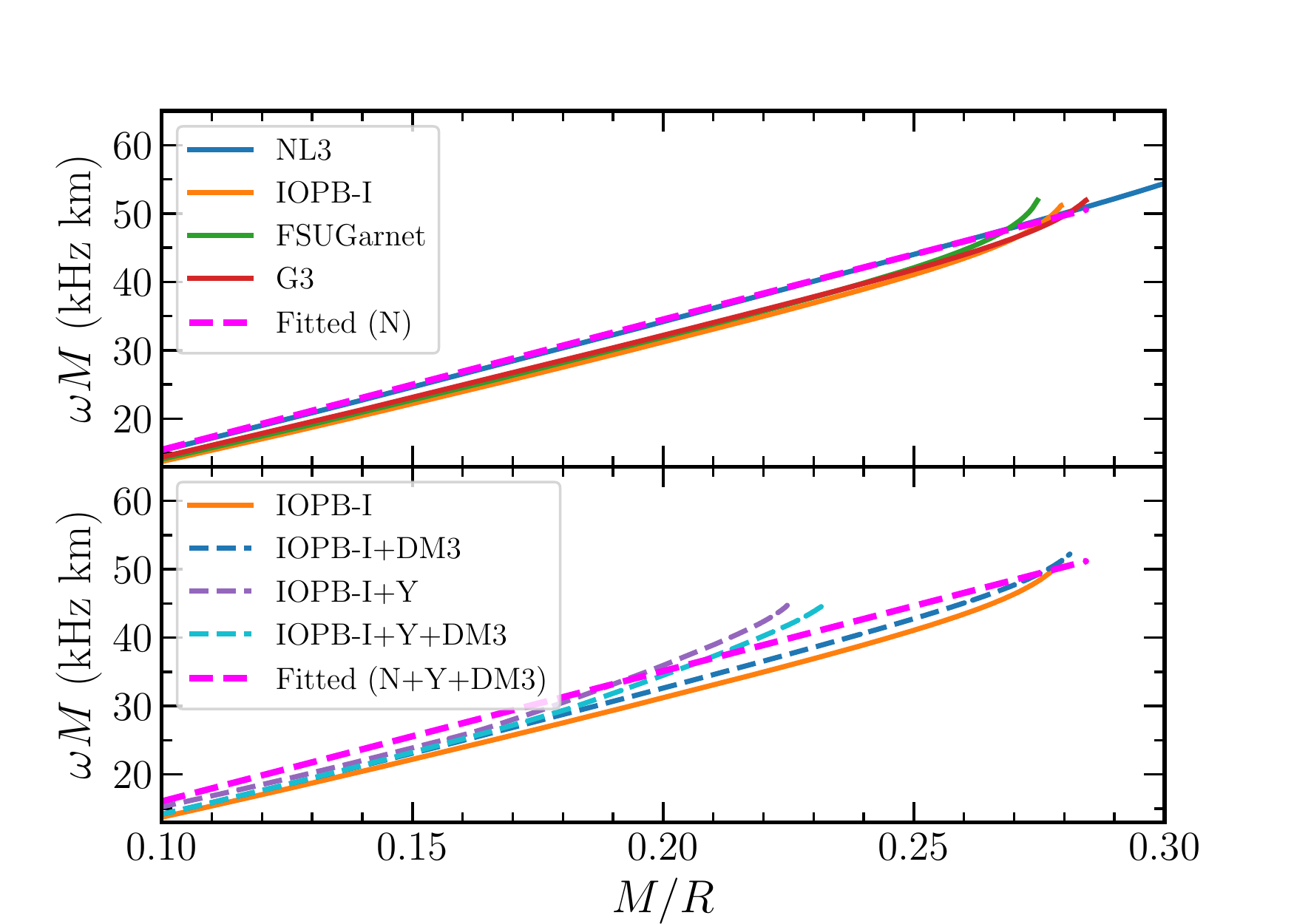}
	\caption{{\it Upper:} Angular frequencies ($\omega=2\pi f$) scaled by mass ($\omega M$) are shown for four different RMF parameter sets. {\it Lower}: $\omega M$ for only IOPB-I with yellow line, IOPB-I+DM3 with blue dashed line, IOPB-I+Y with purple dashed line, IOPB-I+Y+DM with cyan dashed line.}
	\label{fig:fom2}
\end{figure}	
\begin{table}
\caption{Empirical relations, $f(\mathrm{kHz})=a+b\sqrt{\frac{\bar{M}}{\bar{R^3}}}$ between $f$-mode frequency and average density from different works for $l=2$ mode, where $a$ and $b$ are fitting co-efficients.}
\label{tab:table3}
\renewcommand{\tabcolsep}{0.3cm}
\renewcommand{\arraystretch}{1.2}
\scalebox{1.1}{
\begin{tabular}{lll}
\hline\hline
\begin{tabular}[c]{@{}l@{}}Different\\   works\end{tabular} & a(kHz) & b(kHz) \\ \hline
Our fit (N)      & 1.185       & 1.246      \\ \hline
Our fit (N+Y+DM) & 1.256       & 1.311  \\ \hline
AK  fit      & 0.78        & 1.635      \\ \hline
BFG fit      & 0.79        & 1.500      \\ \hline
PC fit       & 1.075       & 1.412      \\ \hline \hline
\end{tabular}}
\end{table}
Latter, these empirical relations were modified by Benhar, Ferrari, and Gualtieri (BFG) ~\cite{Benhar_2004} by using hybrid EOSs. Recently, Pradhan and Chatterjee (PC) ~\cite{Pradhan_2021} have put an empirical relation using hyperonic EOSs. We tabulate the coefficients of the empirical relations from different analyses in Table \ref{tab:table3}. We also find fitting relations (i) for only nucleon EOSs (N) and (ii) for nucleon EOSs with DM/Y and Y+DM (N+Y+DM). The fitting coefficients $a$ and $b$ for different analyses are given in Table \ref{tab:table3} for comparison.

If we measure the $f$-mode frequency of a star, then one can infer the mass and radius of the NS using the fit relations, which can be used to constraint the NS EOS ~\cite{Anderson_1996}. Some works have been tried to find some universal relations between $f$-mode frequency and compactness, or average density ~\cite{Anderson_1996, Anderson_1998, Benhar_2004, Tsui_2005, Lau_2010, Wen_2019}. Thus, the fitting relations are significant to constrain the NS macroscopic properties.  
\subsection{Universal relations}
\begin{figure}
\centering
\includegraphics[width=0.52\textwidth]{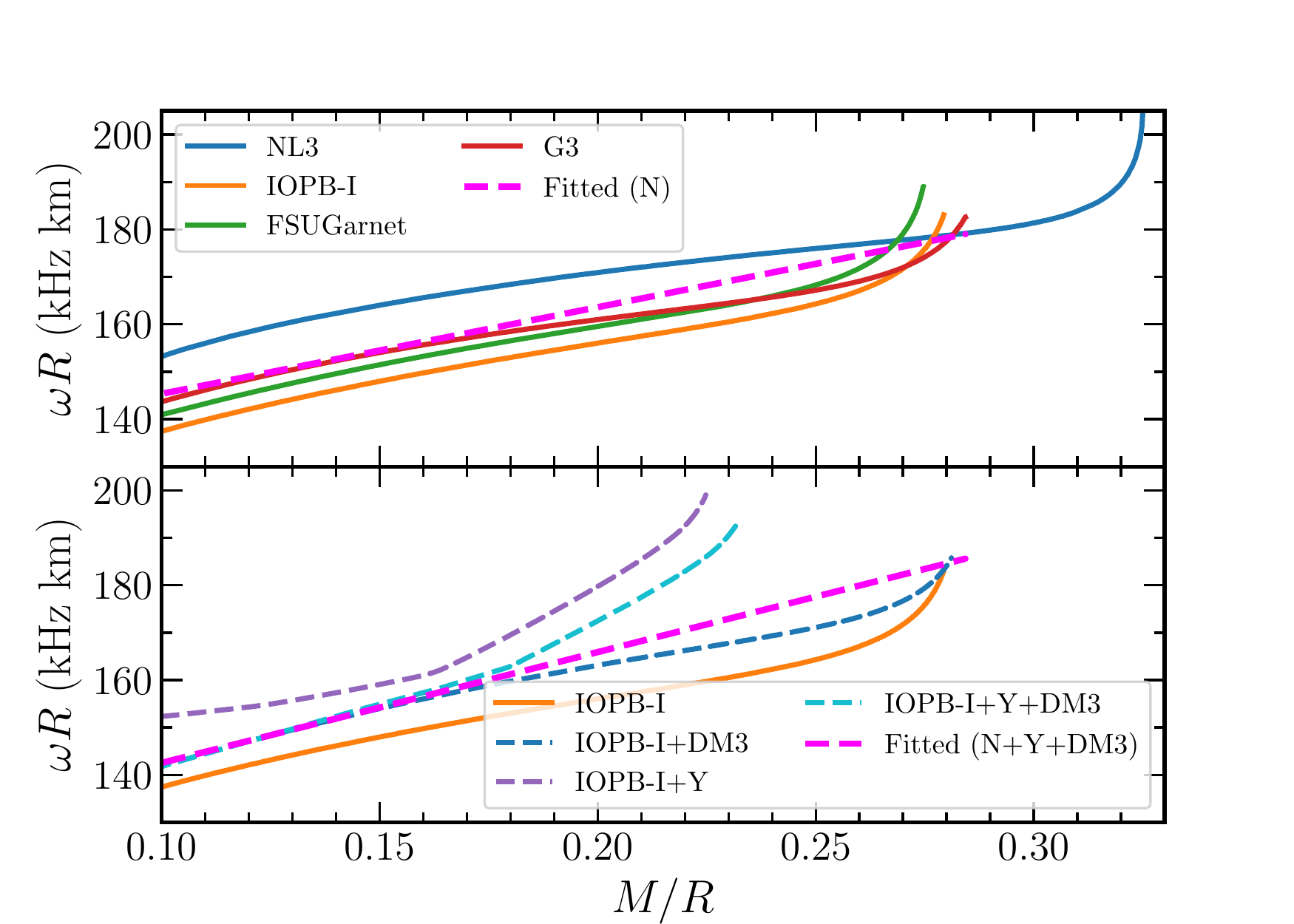}
\caption{Same as Fig. \ref{fig:fom2}, but for $\omega R$ with $M/R$.}
\label{fig:for2}
\end{figure}

Different correlations have been seen between different modes, such as $f$, $p$, and $g$ with compactness/average density. All these relations are quite independent of the EOSs. In Ref. ~\cite{Sotani_2011}, it is observed that the mass scaled angular frequency ($\omega M$) as a function of compactness was found to be universal for $g$-mode frequency. The same type of correlation between $\omega M$ as function of compactness for $p$ and $w$ modes had been proposed in Ref. ~\cite{Salcedo_2014} and for $f$-modes see Ref. ~\cite{Wen_2019}. Thus, we want to check these universal relations as a function of compactness as shown in Fig. \ref{fig:fom2} and Fig. \ref{fig:for2}. We find a universal relation between $\omega M$ and $M/R$. However, the correlation is a little weaker in the case of $\omega R$ with $M/R$. The Universal relations are given as:
\begin{eqnarray}
\omega M (\mathrm{kHz \ km}) &=& a_i\left(\frac{M}{R}\right)-b_i,
\label{eq:uni1}
\\
\omega R (\mathrm{kHz \ km}) &=& c_j\left(\frac{M}{R}\right)+d_j,
\label{eq:uni2}
\end{eqnarray}
where $a_i$, $b_i$, $c_j$ and $d_j$ are the fitting co-efficients in kHz km. The value of $i=1$ for nucleon (N) and $i=2$ for N+Y+DM3. The coefficients are $a_1=190.447$, $b_1=4.538$ and $a_2=190.475$, $b_2=2.984$ for $\omega M\sim M/R$. The values of $c_1=182.585$, $d_1=127.1$ and $c_2=233.596$, $d_2=119.18$ for $\omega R\sim M/R$.
\begin{table*}
\centering
\caption{The observables such as $M_{max}$ ($M_\odot$), $R_{max}$ (km), $R_{1.4}$ (km), $f_{max}$ (kHz), $f_{1.4}$  (kHz), $C_{max}$, $C_{1.4}$, $Z_{s_{max}}$, $Z_{s_{1.4}}$, $\Lambda_{max}$, $\Lambda_{1.4}$ are given for NL3, G3, FSUGarnet, IOPB-I, IOPB-I+DM3, IOPB-I+Y, IOPB-I+Y+DM3 both for maximum mass and canonical NS.}
\label{tab:table4}
\renewcommand{\tabcolsep}{0.3cm}
\renewcommand{\arraystretch}{1.2}
\scalebox{1.1}{
\begin{tabular}{llllllll}
\hline \hline
Observable      & NL3   & G3    & FSUGarnet & IOPB-I & IOPB-I+DM3 & IOPB-I+Y & IOPB-I+Y+DM3 \\ \hline
$M_{max}$       & 2.77  & 1.99   & 2.07  & 2.15  & 2.05  & 1.70  & 1.61  \\ \hline
$R_{max}$       & 13.17 & 10.79  & 11.57 & 11.76 & 11.04 & 11.41 & 10.38 \\ \hline
$R_{1.4}$       & 14.08 & 12.11  & 12.59 & 12.78 & 11.76 & 12.81 & 11.57 \\ \hline
$f_{max}$       & 2.16  & 2.55   & 2.38  & 2.32  & 2.57  & 2.58  & 2.85  \\ \hline
$f_{1.4}$       & 1.78  & 2.06   & 1.94  & 1.87  & 2.14  & 1.92  & 2.22  \\ \hline
$C_{max}$       & 0.31  & 0.27   & 0.26  & 0.27  & 0.27  & 0.22  & 0.23  \\ \hline
$C_{1.4}$       & 0.15  & 0.17   & 0.16  & 0.16  & 0.18  & 0.16  & 0.18  \\ \hline
$Z_{s_{max}}$   & 0.62  & 0.48   & 0.45  & 0.47  & 0.49  & 0.33  & 0.36  \\ \hline
$Z_{s_{1.4}}$   & 0.18  & 0.23   & 0.22  & 0.21  & 0.24  & 0.21  & 0.25  \\ \hline
$\Lambda_{max}$ & 4.48  & 12.12  & 17.67 & 14.78 & 14.39 & 59.58 & 50.58 \\ \hline
$\Lambda_{1.4}$ & 1267.79 & 461.28& 624.81 & 681.27 & 471.06 & 650.55 & 391.68 \\ \hline \hline
\end{tabular}}
\end{table*}
\begin{figure}
\centering
\includegraphics[width=0.52\textwidth]{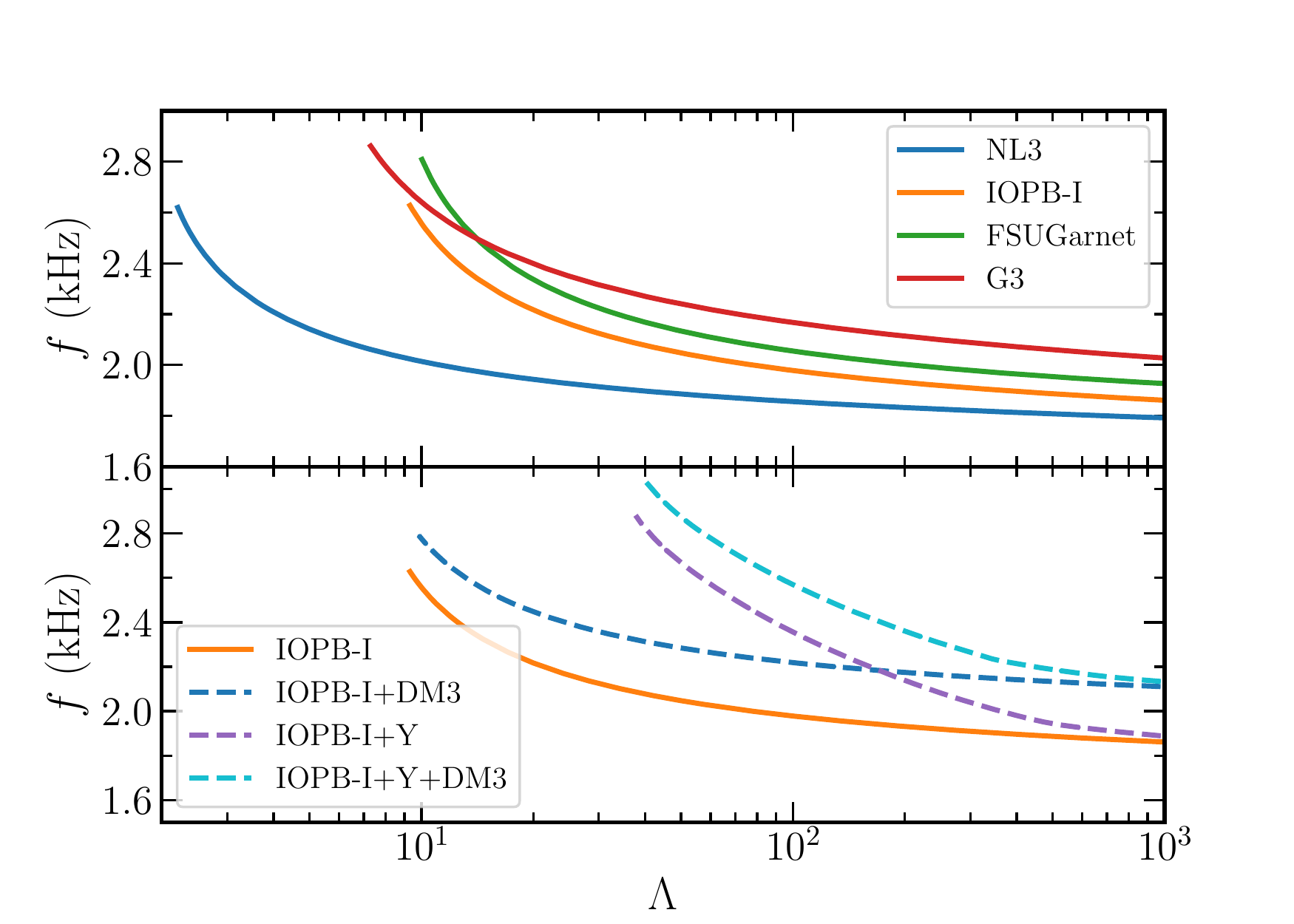}
\caption{{\it Upper:} $f$-mode frequencies are shown with the variation of $\Lambda$ for four different RMF parameter sets. {\it Lower}: $f$-mode frequencies with $\Lambda$ for IOPB-I, IOPB-I+DM3, IOPB-I+Y, and IOPB-I+Y+DM.}
\label{fig:tidal}
\end{figure}

The tidal deformability is the most important quantity, which gives important information about the NS EOSs. The GW170817 event has put a limit on the tidal deformability of the canonical NS, which is used to constraint the EOS of neutron-rich matter at 2--3 times the nuclear saturation densities ~\cite{Abbott_2017, Abbott_2018}. Due to a strong dependence of the tidal deformability with radius ($\Lambda \sim R^5$), it can put stringent constraints on the EOS. Several approaches ~\cite{Bauswein_2017,Annala_2018,Fattoyev_2018,Radice_2018,Mallik_2018,Most_2018,Tews_2018,Nandi_2019,Capano_2020} have been tried to constraint the EOS on the tidal deformability bound given by the GW170817. The dimensionless tidal deformability $\Lambda$ is defined as~\cite{Hinderer_2008,Hinderer_2009, Hinderer_2010}
\begin{eqnarray}
\Lambda = \frac{2}{3}k_2 \ C^{-5},
\label{eq:tidal}
\end{eqnarray}
where $k_2$ is the second Love number, which depends on the internal structure as well as mass and radius of a star ~\cite{Hinderer_2008, Hinderer_2009, Kumartidal_2017, DasBig_2020}. 

We calculate the $f$-mode frequencies as a function of $\Lambda$ for different RMF EOSs, including hyperons and DM. The value of $\Lambda$ decreases when one goes from stiff to soft EOSs and the corresponding $f$-mode frequencies increase contrary to $\Lambda$. The numerical values of $\Lambda_{1.4}$ are given in Table \ref{tab:table4} for different EOSs. The GW170817 event put constraint on $\Lambda_{1.4}=190_{-120}^{+390}$ ~\cite{Abbott_2018}. In Ref. ~\cite{Wen_2019}, they have put a limit on the value of $f$-mode frequencies for 1.4 \ $M_\odot$ are 1.67--2.18 kHz by combining two constraints: the EOS parameter space allowed by terrestrial nuclear experiments and the tidal deformability data from GW170817. Our predicted results of $\Lambda$ are consistent with Wen {\it et al.} \cite{Wen_2019}, except for the case of IOPB-I+Y+DM3. More NS mergers are expected to be measured in the future, which may constraint the $f$-mode frequency more tightly.
\begin{figure}
\centering
\includegraphics[width=0.5\textwidth]{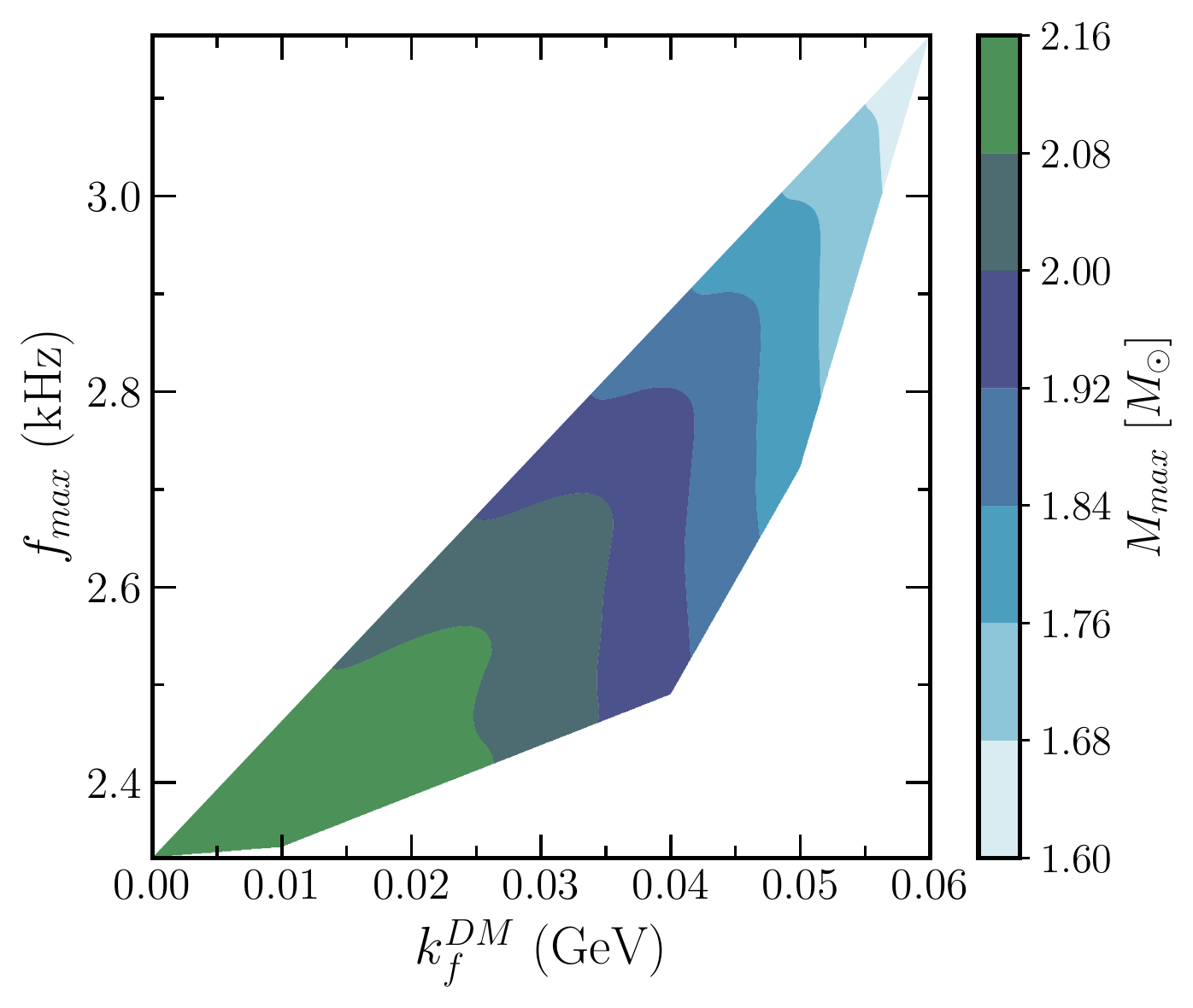}
\caption{$f_{max}$ are shown with different DM momenta for IOPB-I parameter sets. The colour scheme represents the maximum masses correspond to different $k_f^{DM}$. }
\label{fig:fm_dm2}
\end{figure}
\begin{figure*}
\centering
\includegraphics[width=0.54\textwidth]{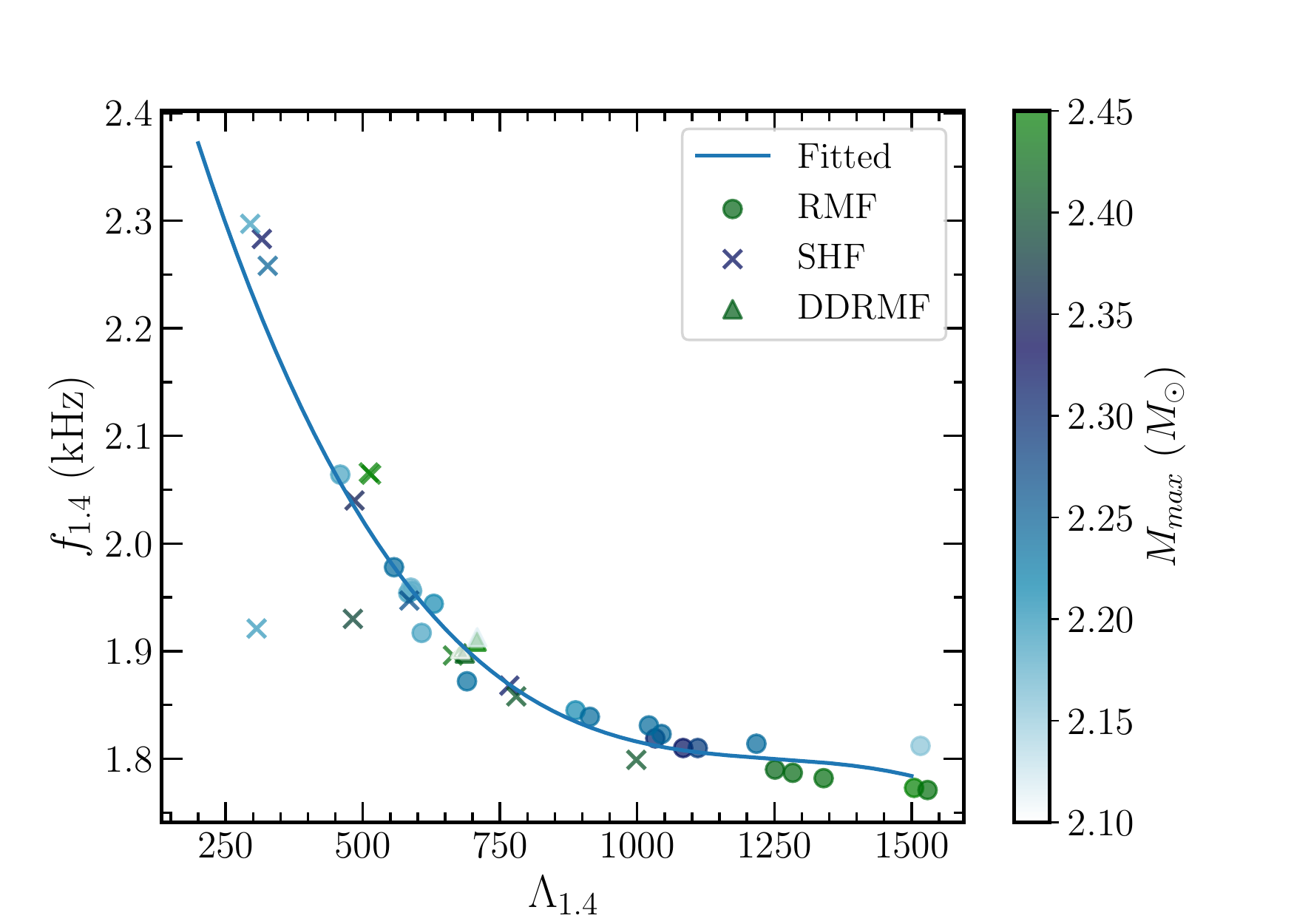}%
\includegraphics[width=0.54\textwidth]{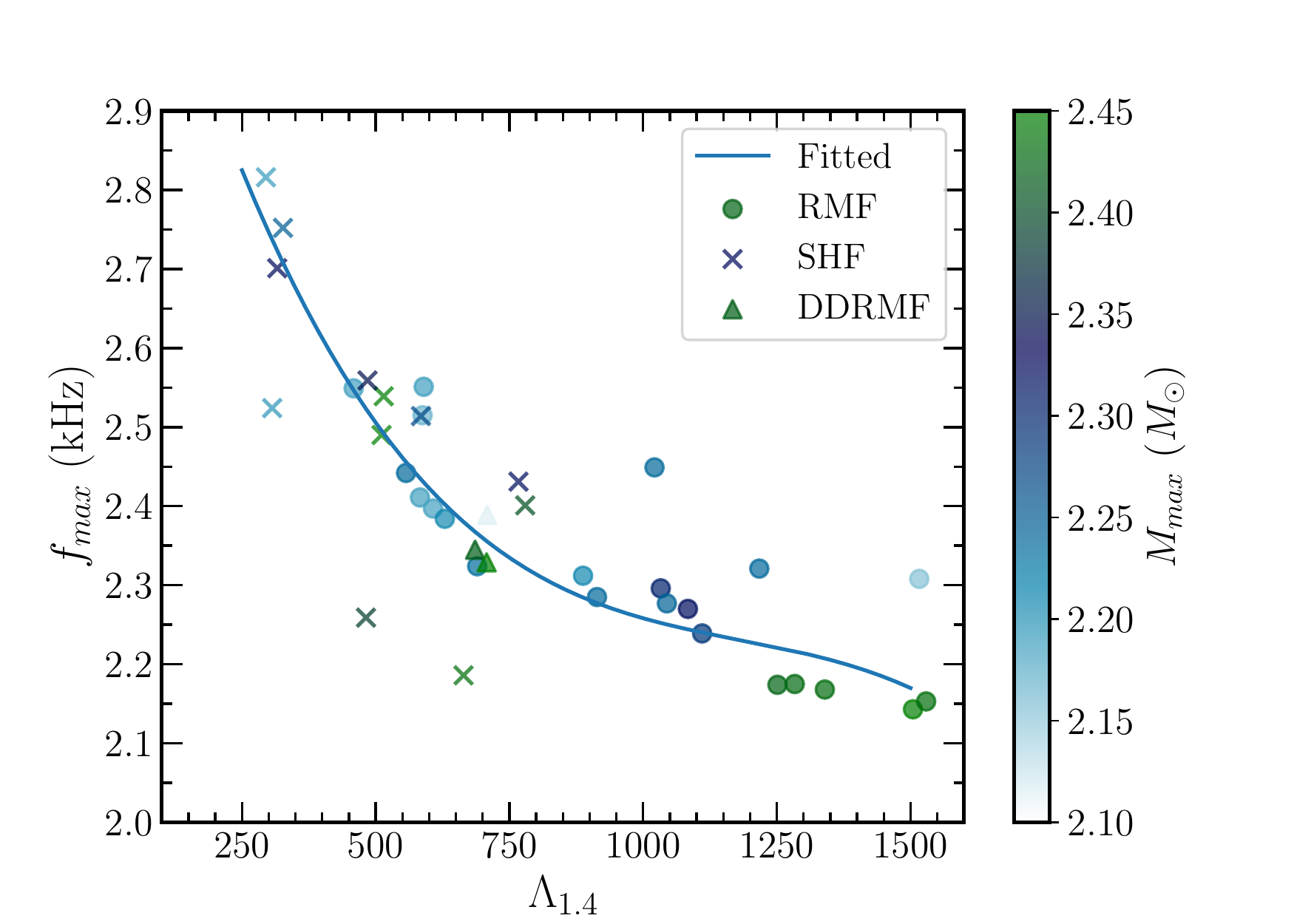}
\caption{{\it Left:} Correlation between $f_{1.4}$ and $\Lambda_{1.4}$ are shown for 41 parameter sets including (RMF, SHF, and DDRMF). The colour bar represents the maximum masses of the corresponding parameter sets. {\it Right:} For $f_{max}$ with $\Lambda_{1.4}$.}
\label{fig:fCLC}
\end{figure*}

In Fig. \ref{fig:fm_dm2}, we plot the values of $f_{max}$ with different DM Fermi momenta. The $f$-mode frequencies increase with increasing $k_f^{DM}$. The range of $f_{max}$ is 2.3--2.4 kHz corresponds to $k_f^{DM}=0.00-0.025$ GeV after that it increases with more proportion beyond $k_f^{DM}=0.03$ GeV. The colour bar represents the maximum masses ($M_{max}$) of the NS for different $k_f^{DM}$. The area of $M_{max}$ at lower $k_f^{DM}$ (green colour) is found to be more as compared to the higher $k_f^{DM}$ and the area slightly reduces from lower $k_f^{DM}$ to higher one. This is because the slight increase of DM percentage reduces the value $M_{max}$ a little bit. When we increases the DM percentage more than 0.03 GeV, the maximum mass reduces a lot. For example, for $k_f^{DM}=$ 0.00, 0.01, and 0.02 GeV, the values of $M_{max}$ are 2.149, 2.146, and 2.119 $M_\odot$ respectively. If we increase the DM Fermi momenta $k_f^{DM}=$ 0.03, 0.04, 0.05, and 0.06 GeV, the values of $M_{max}$ are 2.051, 1.938, 1.788, and 1.614 $M_\odot$ respectively.

We also find some correlations between $f_{1.4}$--$\Lambda_{1.4}$, and  $f_{max}$--$\Lambda_{1.4}$ which are shown in Fig. \ref{fig:fCLC}. This correlation is reported first calculated by Wen {\it et al.} ~\cite{Wen_2019} by taking 23000 phenomenological EOSs, 11 microscopic EOSs, and two quark EOSs. Here, in our calculation, we use 23 RMF EOSs, 14 Skyrme-Hartree-Fock (SHF) forces, and four density-dependent (DDRMF) sets are given in Ref. ~\cite{Biswal_2020} having mass more than $\sim 2 \ M_\odot$. We get a correlation between $f_{1.4}$--$\Lambda_{1.4}$ and  $f_{max}$--$\Lambda_{1.4}$ as shown in figure. The accurately measured tidal deformability of the star from the GW analysis can be used as a constraint on the $f$-mode frequency of a star. We expect that the future GW observation may answer for this type of correlation.
\section{SUMMARY AND CONCLUSIONS}
\label{sec:summ}
Gravitational waves emitted by both the isolated and binary neutron stars are the key quantities that give us enough information about the internal structure and EOSs of the star. If we detect the frequencies of the oscillated star, one can infer the mass and radius of the star from the empirical formula given in the Ref.~\cite{Anderson_1996} with 90\% accuracy. Therefore, the oscillation frequencies are the most promising quantity to constraints the EOS of the NS. The main problem arises to make a Universal empirical relation due to the lack of knowledge on the internal structure of the NS. Several attempts have already been initiated with a little success leaving us an open challenge of the problem. The oscillations frequencies are in the range 1--5 kHz for different modes such as $f$, $p$, $w$, etc. This range of frequencies can't be detected by our present terrestrial detectors, such as LIGO/Virgo. We may be sufficiently advanced to detect that range of frequencies coming from the stars in the future. Prior to this, we have to be equipped with enough possible theoretical solutions to study oscillating stars. Therefore, in this study, we explore the $f$-mode frequencies of the DM admixed NS, and we hope that the results may help the community to explore more about the oscillations frequencies emitted by stars.
	 
In this study, we calculate $f$-mode oscillation frequencies for quadrupole mode within the relativistic Cowling approximations. To compute $f$-mode frequencies, we take RMF EOSs by assuming that the hyperons are present inside the NS. Four well-known RMF forces, such as NL3, G3, IOPB-I, and FSUGarnet, are taken for this calculation. The $f$-mode frequencies decrease with mass for softer EOSs as compared to stiffer ones.  We also include DM as an extra candidate inside the hyperonic star to see their effects on the $f$-mode frequency. The coupling constants for the hyperons-- vectors mesons are calculated using the SU(6) method, while hyperons--scalar mesons are computed by fitting with hyperon potential depth. The baryons-Higgs form factor is assumed to be the same, which is consistent with available theoretical data. The DM-Higgs coupling is constrained with the help of a direct detection experiment and LHC searches.

The $f$-mode frequencies as a function of different astrophysical observables such as $M$, $R$, $\Lambda$, $C$, $Z_s$, etc., are calculated and compared with other theoretical results. The numerical values of $f_{max}$ and $f_{1.4}$ as functions of different quantities are evaluated. The value of $f_{max}$ increases with the addition of either DM/hyperons or DM+hyperons. This is because the EOSs become softer with the addition of DM and hyperons. Softer EOS predicts higher $f$-mode frequency as compared to stiffer one. One can get clear pictures about the variation of $f_{max}$ with different DM Fermi momenta from the Fig. \ref{fig:fm_dm2}. We fit the $f$-mode frequencies with the average density, and the coefficients are in harmony with previous models. The angular frequencies scaled by mass and radius are calculated with compactness. It is found that both $\omega M$ and $\omega R$ follow a linear relationship.

The $f$-mode frequencies with tidal deformability are calculated using the 23 RMF, 14 SHF, and 6 DDRMF equation of states. We find the correlations between $f_{1.4}$--$\Lambda_{1.4}$ and  $f_{max}$--$\Lambda_{1.4}$ are almost consistent with the prediction of Wen {\it et al.} ~\cite{Wen_2019}. The correlation is slightly weaker for $f_{max}$--$\Lambda_{1.4}$ as compared to  $f_{1.4}$--$\Lambda_{1.4}$. Wen {\it et al.} have given a range for $f_{1.4}$ as 1.67--2.18 kHz. Our results correspond to $f_{1.4}$ is in the range 1.78--2.22 kHz for the considered EOSs with DM and hyperons. We hope the discovery of more BNS merger events will open up tight constraints on the $f$-mode frequencies in the future. Thus, it is evident that the effects of either DM/hyperons or DM+hyperons on the $f$-mode frequencies are significant. One can take DM EOSs to calculate $f$-mode frequency in full GR method as done by Lindblom and Detweiler ~\cite{Lindblom_1983}, which may provide more accurate results.    
\section*{Acknowledgement}
HCD like to thank Prof. H. Sotani, Prof. I. F. Ranea Sandoval, and Prof. C. V. Flores for their fruitful discussions about the numerical methods for calculating eigenfrequencies of the star. 
\bibliography{fmode}	
\end{document}